\documentstyle[11pt,aaspp4,epsfig]{article}

\begin{document}

\title {A Variability Study of Pre-Main Sequence Stars in the Extremely 
Young Cluster IC 348}

\author{W. Herbst} 
\affil{Astronomy Department, Wesleyan University, Middletown, CT 06459; wherbst@wesleyan.edu}
\affil{Max-Planck-Institut f\"ur Astronomie, K\"onigstuhl 17, 69117 Heidelberg, Germany} 

\author{J. A. Maley}

\and

\author{E. C. Williams}

\affil{Astronomy Department, Wesleyan University, Middletown, CT 06459} 
\begin{abstract} 

The extremely young cluster IC 348 has been monitored in the 
Cousins I band with a 0.6 m telescope at Wesleyan's Van Vleck 
Observatory. Photometry of 150 stars was obtained on 76 images 
taken on 27 separate nights during the period December, 1998, 
through March, 1999. As expected, spectral characteristics largely 
determine the nature of a star's variability in this cluster. None of the 
stars with H$\alpha$ in absorption were found to be variables. On the 
other hand, all 16 stars identified as CTTS by their H$\alpha$ 
emission equivalent widths and the majority of the 49 WTTS in the 
part of the cluster we monitored showed evidence of variability.  Nineteen stars 
were found to be periodic, with periods ranging from 2.24 to 16.2 
days and masses ranging from 0.35 to 1.1 M$_\odot$.  Seventeen of 
these are WTTS and the other 2 are of unknown spectral class. The 
period distribution is remarkably similar to what is found in the 
Orion Nebula Cluster for stars in the same mass range.  Namely, it is 
bimodal with peaks at 2-3 days and 7-8 days, although there are not 
enough periods known to define these features significantly by the 
IC 348 data alone. The three fastest rotators are also the three most 
massive stars in the periodic sample. It is striking that none of the 
known CTTS were found to be periodic even though they are more highly variable than the WTTS in the cluster. 
This supports the canonical view that WTTS variability is primarily caused by the rotation of a surface with large, cool spots 
whose pattern is often stable for many rotation periods, while CTTS 
variability has an additional component caused by accretion hot 
spots which typically come and go on shorter timescales.  Stars with 
significant infrared excess emission in this sample do tend to be CTTS, while the WTTS (including periodic ones, with one possible exception) show no infrared excess and, therefore, no evidence of disks. Among the CTTS, 
neither H$\alpha$ emission equivalent width nor infrared excess 
emission shows any correlation with degree of variability. 
\end{abstract} 
 
 
\section{Introduction}

It has been known since the pioneering work of Joy (1945) that T Tauri stars are  variable stars. They are now 
recognized to be pre-main sequence (PMS) stars (e.g. Bertout 1989) and their variations are thought to arise from 
a changing pattern of hot accretion zones and cool spots on their 
surfaces (Herbst et al. 1994). Classical T Tauri stars (CTTS), which 
have stronger emission lines and infrared excesses as well as a 
veiling continuum overlying their photospheric spectrum, typically 
vary in irregular fashion on timescales of hours to days with 
amplitudes of a few hudredths to several magnitudes in V. The 
principal source of the modulations is thought to be unsteady 
accretion, which causes the veiling continuum to wax and wane. 
Rotational modulation of the pattern of bright and dark spots on the 
CTTS surface is undoubtedly present, but often masked by changes 
in the spot pattern on timescales comparable to or shorter than a 
typical rotation period of 2 to 20 days. By contrast, the weak T Tauri 
stars (WTTS) are usually periodic variables with V magnitude 
amplitudes of less than 0.75 mag which can be attributed to the 
rotation of a star with very large, cool spots on its surface. These dark 
spots may be associated with the "footprints'' in the 
photosphere of the strong dipole field, which is a central feature of 
the magnetospheric accretion model of TTS (e,g,  Mahdavi and 
Kenyon 1998; Hartmann 1998).

Earlier-type analogs of the CTTS exist, namely, the Herbig Ae/Be 
stars (Herbig 1960; Strom et al. 1972; Hillenbrand et al. 1992) and a 
group of early K through F stars which have no widely used 
moniker. Most of the Herbig Ae/Be stars and the F and G-type TTS 
are variables, with ranges and timescales similar to the CTTS.  As a 
group, these variables have come to be known as UXors, after the 
proto-type UX Ori  (Herbst, 1994; Herbst \& 
Shevchenko 1998; Natta et al. 1999).  No periodicity has been established with 
definiteness for any of these stars (Herbst \& Shevchenko 1998). A 
commonly held view is that the variations result from occultations by 
proto-planets or proto-comets or other dust concentrations in 
circumstellar disks (e,g, Grinin 1994) but Herbst \& Shevchenko 
(1998) propose that variable accretion might, instead, be the cause. 
The most extreme examples of PMS variability, the 
FUors (Herbig, 1977), can change brightness 
by more than 5 magnitudes on timescales as short as a few weeks. 
Unsteady accretion in a luminous disk is widely believed to account 
for their behavior (Hartmann \& Kenyon 1996), although not universally (e.g. Petrov \& Herbig 1992) 

Our ideas about the variations of pre-main sequence stars have 
largely been developed from studies of individual objects that are rarely 
considered within the context of a physically associated group. This 
is partly because the closest and brightest examples tend to be 
members of loose associations, particulary Taurus/Auriga, which are 
too spread out on the sky to be studied {\it en masse}. Early cluster 
studies concentrated on the Orion region where Parenago (1954) 
discovered hundreds of variable stars and Haro, Chavira \& Mendoza (1960) studied the 
flare stars. But photographic plates are poor detectors for work in 
Orion because their blue sensitivity restricts them to spectral regions 
where the nebular light is a serious pollutant and the late-type stars 
are intrinsically faint. Unlike the case for globular clusters, where 
photographic work on variable stars prospered during the era of 
photographic plates (e.g. Hogg 1972), variability studies of extremely 
young clusters languished during the 1960's - 1980's.

The situation has now improved with the widespread availability of 
CCD detectors on small and medium-sized telescopes. By working 
in the far red spectral region, one can avoid the principal nebular 
emission lines and detect stars in Orion almost to the H-burning limit 
with a 0.6m telescope. Variability studies, especially those aimed at 
irregular variables, require lengthy observing runs and cannot usually 
be done on the larger telescopes, for which there is too much shared 
demand. Programs at Van Vleck Observatory on the campus of 
Wesleyan University directed at the Orion Nebula Cluster and, more 
recently, NGC 2264, have been underway since the early 1990's 
(Mandel \& Herbst 1991; Attridge \& Herbst 1992, Eaton, Herbst \& 
Hillenbrand 1995; Choi \& Herbst 1996; Herbst et al. 2000). Other 
recent examples of extremely young cluster monitoring programs 
include those by Adams, Walter \& Wolk (1998) and Stassun et al. (1999), both directed at 
the Orion region, and Makidon et al. (1996) directed at NGC 2264.

The principal focus of the extremely young cluster monitoring 
programs mentioned above has been the discovery of rotation 
periods for PMS stars. These data have been 
used to constrain models of the evolution of stellar angular 
momentum from the PMS to the main sequence phase. However, 
other interesting questions can be addressed by consideration of 
variability more broadly. It should be possible, for example, to test 
the notion that CTTS are more active and less regular variables than 
WTTS because of the role that accretion plays. Variability studies 
could also make a contribution to the study of the clusters 
themselves. For example, detection of irregular (or periodic) 
variations in a star within the cluster field is a good indicator of 
cluster membership which can be used, in conjunction with other 
such indicators as kinematics, location on the HR diagram, emission 
lines, etc. to help isolate true cluster members from field stars. Also, 
the optical variability of PMS stars is almost always neglected (for 
lack of knowledge) when HR diagrams are created. Magnitudes and 
colors are usually based on a small number of measures and could 
be significantly different for a particular star if they had been 
measured on different nights. To quantify the errors associated with 
this practice and to obtain the most applicable data for individual 
stars when possible, a large number of observations are required.

For a variety of reasons, it appears that IC 348 is a young cluster 
uniquely well-suited to a general variability study of its PMS 
population. It is nearby, extremely young and relatively free of the 
nebulosity which complicates photometric studies in, for example, 
the ONC.  The distance is somewhat in doubt since the most 
straightforward and precise method - averaging parallaxes of cluster 
members measured by Hippacrcos - leads to a result (260 $\pm$ 25 
pc; Scholz et al. 1999) which is marginally closer than traditional 
main sequence fitting methods (316 pc; Herbig 1998, hereinafter 
H98). The uncertainty in the distance translates to an uncertainty in 
the age of the PMS stars because it is determined by fitting their 
luminosities to models. At the larger, photometric distance, the 
median age of the PMS stars is about 1.3 million years, whereas at 
the Hipparcos distance, the stars are closer to 3 million years old, 
according to the models of D'Antona \& Mazzitelli (1994).  Since 
the typical age of a PMS star in the ONC is about 0.8 My according 
to the same models (Hillenbrand 1997), IC 348 is between 1.5 and 4 
times older than the ONC. Of course, there is an age range in both 
clusters, so individual stellar ages determined by location on the HR 
diagram overlap with each other.  The cluster has about 60 M$_{\odot}$ in stars with an average mass of about 0.5 M$_{\odot}$ within 
its "core'' radius of 4\arcmin (H98). This means that there are more than 
one hundred late-type members which can be monitored for 
variability within a 10$\arcmin$\ field.  The earliest type star 
(BD+31$\deg$ 632) is of spectral class B5, so it illuminates only a 
faint reflection nebula, not a bright emission nebula. The only 
disadvantage for photometry is that the cluster is rather heavily 
embedded in the local dust cloud out of which it presumably formed, 
since BD+31$\deg$ 632 is insufficiently luminous to have cleared 
the region of grains. This means that extinction is rather high and 
variable from star to star. It restricts monitoring programs with small 
telescopes to a far red (e.g. Cousins I) band.

A significant attraction of IC 348 for a variability study is that it has 
been the subject of several recent, detailed investigations by a variety 
of techniques, which have characterized the visible stellar population 
in important ways. The early history of this work is recounted in 
most of the papers below; here we mention only the contributions of 
the last five years, beginning with the near-infrared studies of Lada 
\& Lada (1995) and Luhman et al. (1998). These have provided a 
deep stellar census and spectral types for most of the stars in the 
core region of the cluster. Information on the presence or absence of 
disks comes from the near-IR colors as well as veiling estimates in 
the IR spectroscopy and Br$\alpha$ and H$\alpha$ emission 
line strengths. A detailed optical spectroscopic and photometric 
study by Herbig (H98) provides the data to assess extinction and 
place the stars on an HR diagram, from which masses and ages can 
be inferred. This is supplemented by the photometric study of 
Trullols \& Jordi (1997; TJ). An H$\alpha$ survey reported in H98 
provides information on whether stars are CTTS, WTTS or show 
absorption lines. A proper motion study by Scholz et al. (1999) has 
isolated the motion of the cluster from that of nearby and distant field 
stars and provided membership probabilities for some stars in the 
core. An X-ray study of the cluster by Preibisch et al. (1996) 
provides ROSAT data on many core members. Finally, Duchene, 
Bouvier \& Simon (1999) have searched for optical binaries in the 
cluster using adaptive optics techniques. A variability study would 
seem to be a valuable addition to this outburst of activity on an 
important, nearby cluster. 

\section{Observations and Initial Reductions}

The observations were obtained during during the period December, 
1998, to March, 1999, with a 1024x1024 Photometrics CCD 
attached to the f/13.5, 0.6 m Perkin telescope at Van Vleck 
Observatory on the campus of Wesleyan University. The size of a 
pixel is 0.6$\arcsec$, so the size of the field is 10.2$\arcmin$.  All 
data were obtained through a Cousins I filter. A sequence of five 
one-minute exposures on the selected cluster field, shown in Figure 
1, was taken each clear night, and on most nights the sequence was 
repeated, often more than once, after a couple of hours. Each image 
in a sequence was shifted and combined with the others to form a 
single image with an effective exposure time of five minutes and an 
expanded dynamic range. Since we do not have an autoguider on 
this telescope, a single five-minute exposure would often have 
resulted in trailed images. The sequence of shorter exposures which 
is shifted and summed serves as a "software autoguider'' which has 
the added advantage of expanding the dynamic range of the data. 
Read-out times for the chip are only 8 seconds, so this procedure 
does not compromise observing efficiency much. Flat-fielding was 
accomplished using twilight flats which were obtained each night. 
Bias frames and dark frames were also obtained each night and 
appropriate corrections applied to the images using standard IRAF 
tasks. A log of the observations is given in Table 1. Altogether, we 
obtained 76 images with an effective integration time of 5 minutes 
each on 27 separate nights.

The seeing, as measured by the full-width at half-maximum of the 
stellar profiles, ranged from $\sim$1.5$\arcsec$ to  
$\sim$4$\arcsec$ with a mode of 2.5$\arcsec$. An image with 
excellent seeing, displayed in Figure 1, was chosen as the reference 
to which all others were shifted and from which we constructed our 
catalog of stars for photometry. This was done by careful visual 
examination of the frame and selection of all objects which were 
"obviously'' stars.  A total of 151 objects was selected, one of which 
turned out not to be visible on any other image and is almost 
certainly not a star. The positions of these objects on our reference 
image were transformed to J2000 coordinates by cross-identifying 
some with stars in Herbig's (H98) list and using a FORTRAN 
program kindly provided by G. Herbig to accomplish the 
transformation. Comparison with the positions of 25 stars which we 
observed in common with Scholz et al.'s (1999) proper motion study 
revealed a mean difference in our positions in right ascension and 
declination as, respectively, $-0.5\arcsec$ and $+0.8\arcsec$. 
Positions for our program stars and cross-identifications with H98 and TJ are given in Table 2. 

All images were aligned with the reference image using the 
IMALIGN procedure in IRAF and then aperture photmetry was 
performed with the APPHOT procedure. An aperture radius of 7 
pixels (4.2$\arcsec$) was used. The sky was determined as the 
median in an annulus of inner radius 15 pixels and outer radius 25 
pixels after rejection of points in excess of 3 sigma from the mean. 
Because we selected stars on the very best image and went 
approximately to the limit of that image, there are some stars which 
could not be measured photometrically on nights of poorer seeing, 
either because of small signal to noise ratio or confusion with a close 
companion. The number of images on which we obtained a 
photometric result from APPHOT is given for each star in Table 2. 

A critical step in the reductions is to identify a set of comparison 
stars whose average magnitude will define the reference brightness 
on each night. This was done by a "bootstrap'' technique, as follows. 
First, we used all the stars which had a measurement on every night to 
define an initial reference magnitude. Then, we excluded those stars 
with large scatter and re-computed the reference magnitude as the 
mean of the remaining stars. This process was repeated several times 
until the potential comparison objects had been whittled down to a 
set of 6 stars. The scatter of these 6 relative to the reference 
magnitude was characterized by an average sigma of only 0.007 
magnitudes, which we judged to be suitably small, and they were 
adopted as the comparison stars.  They are stars 11, 12, 17, 24, 26 
and 27.  It is important to note that this process of selecting 
comparison stars was done without reference to their spectral or 
other characteristics. We did not, for example, deliberately choose 
non-members of the cluster, or early-type members because {\it a 
priori} they might be expected to vary less than late-type cluster 
members. In fact, we discovered at the end of the analysis that 2 of 
the 6 selected comparison stars (nos. 12 and 26) were WTTS with 
low amplitude, periodic variations.  The averaging procedure reduced 
the effect of their variations on the reference magnitude to nearly 
undetectable levels. Only one of the remaining four stars, number 17, 
revealed an effect of the contamination - a significant peak in its 
periodogram at the same period (2.24 days) as star 12, but 
180$\deg$ out of phase and with less than one-sixth of the 
amplitude (i.e. $\sim 0.01$ mag). In our view, this level of 
contamination was so small that it was not worth re-doing the 
photometry using {\it a posteriori} knowledge, especially since it 
would have cast some doubt on a principal result, namely, the 
relationship between spectral characteristics and variability. 
Differential magnitudes relative to the average of the selected 
comparison stars were formed from our APPHOT data and these 
constitute the time series that we analyze in this paper. 

An important factor influencing our photometric results for some 
stars is the presence of a nearby companion. Pairs with 
separations of less than 8$\arcsec$ (or larger if one star is very bright) can have an effect on the 
photometry in two ways. First, close binaries will be measured as 
single stars, elevating the brightness of each to a common value. 
Second, variable seeing can cause different amounts of light from a 
companion to be included within the aperture of the program star on 
different nights, simulating variability. In principle, one can 
compensate for these effects by using a profile-fitting photometric 
technique. In practice, however, we found that the increased 
complexity associated with this is not worth the effort for our 
relatively uncrowded fields. Instead, we list in Table 3 all of the 
known visual binary stars in our field and we take their status into 
account when analyzing the results. 

\section{Transformation to a Standard System}

Since our data were obtained with a CCD and filter combination 
chosen to match the Cousins I band (Bessell 1990), a linear 
transformation from instrumental magnitude (i) to standard 
magnitude (I) may be expected. We attempted to match the local 
photometric systems of Trullols and Jordi (TJ) and Herbig (H98), as 
shown in Figure 2, both of which were tied to the Cousins system by 
observation of standard stars from the list of Landolt (1992). Boxed 
symbols on these figures are stars with close companions from Table 
3, which may well have contaminated photometry. It is clear that 
they lie above the general i-I distribution, especially in the figure 
based on the H98 data, indicating that they are measured as too 
bright by us.  It is also obvious that a large systematic difference 
exists between the observations of TJ and H98, as has been noted by 
H98. For example, TJ have no stars redder than R-I = 2.0, whereas 
H98 has many. 

The systematic difference in R-I is clearly shown in Figure 3 where 
we plot the TJ values against the H98 values for stars in our sample.  
For R-I $<$ 1.5 it is evident that the agreement is reasonable, but for 
redder stars there is a huge scatter and essentially no correlation. 
Returning to Figure 2, it is clear that our data are much more in 
accord with the results reported by H98. In the bottom panel we see 
that, ignoring the contaminated stars, there is a very nearly linear 
relation between i-I and R-I. A least squares fit to the uncontaminated data (excluding star 13, see below) is shown. It is interesting that the exact same fit, as shown in the top panel of Figure 2, matches quite 
well with the narrow band in the TJ data defined by most of the stars 
with R-I $<$ 1.5. In fact, the scatter around the line for R-I  $<$ 1.5 in the top panel (TJ data) is smaller than for the H98 data. 
This is probably because these stars are 
the earlier-type members of the cluster which are both brighter and, 
as we show below, have less intrinsic variability. However, redder 
stars have obviously been assigned colors by TJ which are much too 
blue, resulting in a truncated color distribution and large scatter in i-I 
between 1.0$<$R-I$<$2.0. Our conclusion is that the color systems 
match for the bright stars with R-I$<$1.5 but that the TJ 
transformation is invalid for redder stars.  The form of the 
discrepancy suggests that TJ may have extrapolated a quadratic fit 
beyond their reddest standard in deriving colors for the redder 
program stars.

In accordance with the discussion above, we transformed our 
instrumental magnitudes to Cousins I magnitudes using the 
following relationship derived for stars in common with H98: $$I = i + 12.074 + 0.099 \times (R-I)$$ 
R-I values from H98 were used for the transformation when 
available. For the brighter stars, which were not measured by Herbig, 
we used the colors from TJ, which is valid since they all have R-I 
$<$ 1.0. For the fainter stars not measured by H98 we have assumed 
R-I = 2.0 in transforming to the standard system. Of course this is not 
ideal but is adequate for our purposes. Average I magnitudes are 
given for all of our stars in Table 2. The only surprising result is for Star 13, which is a G0 star with no indication of variability in our data nor in H98 or TJ. Our result, I = 11.99, is in good agreement with TJ, who find I = 12.02, but differs substantially from H98, whose data indicate I = 12.60. It is uncertain at present whether this is an actual variable (possibly an eclipsing binary or more likely an UXor) or whether the H98 measurement is in error. An additional season or two of monitoring may resolve the question; here we do not regard the star as a definite variable.

\section{Variability}

\subsection{Scatter in the Data as a Function of Brightness}

We use the standard error ($\sigma$) of our measures to address the 
issue of variability. The objective of the analysis described in this 
section is to remove the contributions to $\sigma$ from random and 
systematic photometric errors leaving only that part which may 
reasonably be assigned to intrinsic variability. We begin by 
displaying, in Figure 4, log$_{10}$ $\sigma$ as a function of I 
magnitude. As in Figure 2, boxed points are stars with companions, 
(c.f. Table 3) whose photometry is possibly contaminated. In 
this case, the problem with contamination is that it can vary from 
night to night on account of changes in seeing, which simulates 
stellar variability.  The uncontaminated stars have a distribution with 
a well-defined lower envelope which we assume, for the moment, to 
define the locus of non-variable stars.  This assumption will be 
justified in what follows.
 
It is clear from Figure 4 that there are two regimes to the data, as one 
would expect on general grounds. For the brighter stars (I$<$13.5) 
the lower limit to $\sigma$ is essentially a constant, with the value of 
0.009 mag. We interpret this as a limit set by systematic effects - 
probably coming mostly from flat-fielding errors and residual 
variability in the reference magnitude - to our photometric accuracy 
for even bright stars. The lower limit to $\sigma$ for stars fainter than 
I = 13.5 is well represented by a straight line of slope 0.30. This is 
steeper than what would be expected if random errors with a Poisson 
distribution were the sole cause of the noise, in which case the slope 
would be 0.20. Also, there is some evidence that the slope increases 
with decreasing brightness. Presumably these facts reflect the 
influence of sky measurement errors on the photometry, although we 
have not attempted to prove this. For our purposes it is sufficient to 
represent the random and systematic errors of the photometry 
$\sigma_{err}$ as follows: $$\sigma_{err}=0.009\ (for\ I  \le 13.5)$$
$$\sigma_{err}=0.009\ \times\ 10^{0.30(I-13.5)}\ (for\ I >13.5)$$
The lines defined by these equations are shown on Figure 4.

\subsection{Variability as a Function of Spectral Characterisitics}

Figure 5 is a $\sigma$ vs. I plot for the subset of the program stars with 
spectral data. Close pairs, from Table 3, whose photometry is 
possibly contaminated by a companion and stars fainter than I $\sim$ 17 
mag, where random errors are very large, are not displayed.  Note that 
boxed symbols on this figure are periodic variables (see below), not 
contaminated stars. Various symbols indicate (primarily spectral) 
categories defined by the criteria given in the notes to Table 2. For the purposes 
of this figure, we have combined the E (early-type cluster members), 
N (non-members) and G (G and early K-type members without 
emission lines) categories, displaying them with a common symbol 
(diamond). These stars, or at least the E and N categories, are not 
expected to be variable stars. It is gratifying to find that, in fact, they 
define the lower evnvelope of the $\sigma$ vs. I distribution quite 
well. The fit to this envelope derived above is shown and is obviously 
a good representation of the  relationship for the expected non-variable stars with I $<$ 17.  By contrast, the CTTS and WTTS are 
found mostly above the line, indicating significant variability.

Three principal results of this study, which happen to confirm 
expectation based on examples studied in isolation (cf. Herbst et al. 
1994), are evident upon examination of Figure 5. First, whereas the 
absorption line stars show no evidence of variability, most of the 
CTTS and WTTS are clearly variable stars.  
Second, the CTTS in this cluster, as a group, are more variable than 
the WTTS. While there are a couple of CTTS which showed very 
little variation during the time interval of this work, most of them 
were clearly variable, and several by amounts exceeding 1 
magnitude. By contrast, a significant fraction of the WTTS had 
variations which were too small to detect. Third, and perhaps most 
interesting, is the fact that not one of the stars found to be periodic is 
a known CTTS, whereas 17 of the 19 are known to be WTTS.  This 
cannot be a selection effect, since it would be quite easy to find 
periods for the large amplitude, relatively bright CTTS in IC 348, if 
they were periodic during the course of our study. Instead, it must 
reflect the fact that the variations of the CTTS are qualitatively 
different from the WTTS in a way which supports the canonical view 
of the cause of the variations, as expressed in the first paragraph of 
this paper. We emphasize that no reference to spectral characteristics 
was made until all facets of the variability study, including the 
periodogram analysis, were completed.

It is evident from the discussion above that the spectral feature which 
matters most in predicting variability is the one which separates TTS 
from absorption line stars and CTTS from WTTS, namely,  
H$\alpha$ emission. To further elucidate its importance we display, 
in Figure 6, a variability parameter, $\sigma_{var}$, as a function of 
the equivalent width of H$\alpha$ as determined by H98 or Luhman 
et al. (1998). The variability parameter is defined as the portion of the 
total ($\sigma$) above that which can be attributed to photometric 
errors, i.e. $$\sigma_{var} = \sqrt{\sigma^2 - \sigma_{err}^2}$$
Some of what we attribute to real variation by this definition could 
obviously come from influences such as variations in sky brightness 
caused by reflection nebulosity or accidental errors of various sorts, 
but we assume here that the quantity basically measures intrinsic 
variability. Obviously, the error in $\sigma_{var}$ increases 
substantially with decreasing brightness, and it is pointless to discuss 
the quantity for stars fainter than I = 17.

It is interesting to note on Figure 6 that, while the CTTS are 
generally more variable than the WTTS, there is no correlation 
within either group between equivalent width of H$\alpha$ and 
degree of variability. So, while H$\alpha$ equivalent width works, in 
a general way, to predict variability it is obviously not the only factor. 
Perhaps part of the scatter in the figure is caused by the non-simultaneity of spectroscopic and photometric observations and the 
substantial variability in H$\alpha$ displayed by the more active 
CTTS. Measurements reported by H98 and by Luhman et al. (1998) 
for the same CTTS at different epochs can vary by large amounts. 
Herbst et al. (1994) argued that there was generally a good 
correlation between H$\alpha$ flux and brightness in CTTS. For 
highly variable objects, especially when observed at widely different 
epochs, one might expect to find a large scatter on such a plot. It 
should also be noted that, as in the general population of brighter 
CTTS, there are some stars which have rather small photometric 
variations. These may be objects in which the accretion is rather 
steady on timescales of days, months and even years, or which were 
in a fairly low accretion state at the time of the photometric 
observations. It would obviously be interesting to know what the 
range of H$\alpha$ variability was for such stars and, ideally, to have 
simultaneous photometric and spectroscopic observations. Turning 
the variability argument around, the division between CTTS and 
WTTS in this cluster, based on their photometric properties, seems to 
be at an equivalent width of 11 \AA, essentially the canonical 
value.  We turn now to a more detailed look at the variations of, first, 
the irregular variables and, then, the periodic stars.

\subsection{Irregular Variables}

Light curves for four of the more extreme examples of CTTS 
variability are shown in Figure 7. Star 23 is the most highly variable 
object in our study, having an I magnitude range of about 2 
magnitudes. This makes it comparable in activity to the largest 
amplitude CTTS found in the general population (Herbst et al. 1994). 
The star is of typical spectral type for a CTTS (K6). However, its 
light curve is more similar to the UXors (Herbst \& Shevchenko 
1998) in that it is punctuated by deep minima, of which two are 
visible in Figure 6. Rather than being a typical CTTS this may be a 
transition object between the GTTS, such as RY Tau and CO Ori, 
and the CTTS. It is more similar to the CTTS in the respect that its 
H$\alpha$ line is strongly in emission.

Stars 59 and 141 are the next highest amplitude CTTS and show 
somewhat similar light curve forms. Both tend to be near the brighter 
portions of their ranges but display occassional deep minima (one 
each). In the case of Star 59, this may have been a short-lived event, 
since the upward transition was quite abrupt. In the case of Star 141 
there seems to have been a steady transition to a minimum which 
occurred over about a two week period. If one accepts the hypothesis 
that accretion luminosity is principally powering these stars in their 
bright states, then the minima may be the brightness levels of the 
stars which correspond most closely to their photospheric 
luminosities. As such, they would represent the magnitudes which 
ought to be assigned to the stars when they are placed on an HR 
diagram to derive masses and ages. Alternatively, it is possible that 
the deep minima are caused by occultation events. Bouvier et al. 
(1999) have recently proposed an occultation model for the extreme 
CTTS AA Tau. 

Star 65 is rather interesting and unusual. Its light curve shows 
relatively little (although some) scatter on short timescales, but 
considerable variation over the several months that we monitored it. 
The star appeared in our periodogram analysis with a strong signal at 
a period of about 60 days, and a glance at the light curve shows why. 
There are evidently about 1.5 "cycles'' of what could be a periodic 
change. However, with so small a number of cycles it is impossible 
to conclude that the star is periodic, and it would be difficult to know 
how to interpret such a long period. Obviously, another season or 
more of monitoring is necessary to reach a firm conclusion, but we 
suspect that this is merely an irregular variable whose timescale for 
variation tends to be much longer than is typical for a CTTS.

By contrast with the CTTS, most of the larger amplitude WTTS are 
periodic variables, and we discuss them in the next section. In Figure 
8, we display the light curves of the four most active WTTS which are 
not periodic. We note that the range of their activity is limited to 
about 0.5 magnitudes - smaller than is the case for the largest 
amplitude CTTS and consistent with what is found for WTTS in the 
general population. The most active non-periodic WTTS is star 56 = 
H214, which is also number 75 in the list of Luhman et al. (1998). It 
was classified as K8 by H98, but an infrared spectrum suggests M2-M4 (Luhman et al. 1998). These authors also note that it may have 
infrared veiling at the level of 50\% and enhanced CO absorption at 
the level of 25\%-50\%. Herbig noted that it was a variable star. 
Neither previous study speaks to the presence or strength of 
Hydrogen emission lines from which we inferred that they must be 
weak; hence the classification here as a WTTS. Perhaps the star is, 
actually a CTTS, or is a transition object. Its light curve is also 
characterized by the presence of occasional minima superimposed 
on a smaller variation around maximum light. 

The next three largest non-periodic WTTS with I $<$ 16 are stars 
101 (= H78), 87 (=H174) and 47 (=H170) which are early M stars 
with H$\alpha$ equivalent widths of 8, 10 and 7 angstroms 
respectively. These are at the high end of the WTTS distribution and 
it seems possible that the variability characteristics of these stars also 
reflect a transition from CTTS to WTTS. A small amount of irregular 
variability caused by some lingering accretion events could easily 
add enough "noise'' to the spot variations to make it impossible for 
us to detect the rotation period in one season. Alternatively, the cool 
spot patterns on these stars may simply not have been sufficiently 
stable during the past season to allow determination of a period. 
When observations of extremely young clusters such as the Orion 
Nebula Cluster (e.g. Herbst et al. 2000) are obtained over many 
years, we often find that objects which were not periodic one year, 
appear periodic the next. It will be interesting to see as we continue 
our monitoring of IC 348 if that turns out to be the case for these 
stars.

\section{Periodic Variables}

\subsection{Detecting Periodic Variations}

The method used to identify periodic variables here is similar to that 
described by Herbst et al. (2000). Briefly, periodograms are formed 
from the I magnitude time series for each star. The method of Scargle 
(1982), as formulated by Horne \& Baliunas (1986), is used to do this. 
Every star is included in this search, regardless of whether it 
showed a significantly non-zero value of  $\sigma_{var}$.
The periodograms are computed for the entire realm of  plausible or 
possibly detectable frequencies, from $\sim$5 day$^{-1}$ to 1/120 
day$^{-1}$. The Sampling Theorem does not inhibit us from 
detecting high frequency variations because our data are, of 
necessity, irregularly spaced. In other words, having several 
observations per night on many nights allows us to discover periods 
as short as a few hours, much less than the quasi-Nyquist period. 
However, with irregularly spaced data such as these one must be 
cautious about calculating a false alarm probability (FAP) based on 
the power levels of the periodograms. In particular, the 
approximation of Horne \& Baliunas (1986), which is valid for 
uncorrelated data points, cannot be used (cf. Eaton, Herbst \& 
Hillenbrand 1995; Herbst \& Wittenmyer 1996). 

Since there is no rigorous procedure for calculating a FAP, but we 
wish to have a quantitative method, nonetheless, we proceed as 
follows. Two periodograms were constructed for each star, one based 
on the entire time series and the other based on a modified series in 
which all the data obtained on a single night were averaged to define 
a single point. The power levels for the averaged time series are 
much lower and it is reasonable to use the FAP formulation of Horne 
\& Baliunas (1986) to assess their significance since the data on 
different nights are not so obviously correlated as in the original time 
series.  Twenty-one stars were found with at least one peak less than 
or equal to the limit of FAP = 0.012 (slightly extended to 
accommodate star 73) which we use to identify periodic variables. 
(Two stars, 17 and 65, are rejected for reasons discussed below.) This 
exercise provided guidance to us on how to interpret the power 
levels of the unaveraged time series. In Table 4 we list the stars which 
we identify as having significant periodicity. The FAP determined from the Horne \& Baliunas (1986) 
formula applied to the time-averaged periodograms is also given. 
Only when the FAP was close to the limit of 0.01 was there any 
uncertainty about whether to include the star in our list of periodic 
variables. A sample of periodograms covering the range of the data 
in two parameters is shown in Figure 9. Included in this are the stars 
identified as periodic with the highest and lowest powers (stars 41 
and 49, respectively), and with the longest and shortest periods (stars 
53 and 12, respectively). Note that the secondary peaks which are 
always present in these periodograms at short period are "beat 
frequencies'' between the star's rotation period and the natural 
sampling interval (1 day$^{-1}$) imposed by the rotation of the 
earth and the fixed longitude of VVO.

Admittedly, the procedure employed here in assessing the FAP is 
not mathematically rigorous, but that is the nature of the problem 
when dealing with irregularly spaced observations. Experience with 
this method has led to highly reproducible results in the Orion 
Nebula cluster (see Stassun et al. 1999 and Herbst et al. 2000). We 
regard the phased light curves of the variables as a more reliable 
indicator ({\it albeit} qualitative) than the FAP of the significance of 
the detected periodicity. For this reason we display, in Figures 10 to 
12, the light curves of the 19 stars identified as periodic. They are 
characteristic of PMS spotted variables in all respects. The 
amplitudes range from a few hundredths to a few tenths of a 
magnitude, except for star 73 (shown on a different scale) which has 
a range of 0.7 mag. The light curve shapes are not exactly sinusoidal 
(or any other consistent shape), but tend to be continuously variable, 
i.e. without phases of constant brightness. The period range is 
precisely what is found for the ONC (Herbst et al. 2000) and, as we 
show below, the frequency distribution of periods is also similar. 

Before discussing the periodic stars further, we mention two cases of 
stars with high power levels in their periodograms which do not 
appear in Figures 10 to 12 and one which appears twice. Star 65 has 
a highly significant peak in its periodogram at a period of 73.4 days. 
It can hardly be regarded as periodic, however, since less than 2 
cycles are contained within our observing period. The light curve of 
the star was displayed in Figure 7, where it may be seen why a peak 
in the periodogram occurs. If this star is, indeed, periodic with such a 
long timescale it will require at least one more year of monitoring to 
prove that. Here we regard it as likely to be an irregular variable on a 
typical timescale of months and its apparent periodicity as accidental. 

Star 17 (a comparison star included in the reference magnitude) 
showed a very small amplitude variation ($\sim 0.01$ mag) but with 
a significant power level at a period of 2.24 days.  This is exactly the 
same period as Star 12 (another reference star), and it is precisely 
180$\deg$ out of phase with it.  Clearly, what has happened in 
this case is that the actual periodic variation of  Star 12, at a level of 
about $\pm 0.04$ mag, has been diluted by averaging with the other 
comparison stars, but not entirely eliminated from the comparison 
magnitude. The highly stable Star 17 reflects this at the level of $\pm 
0.01$ mag. In principle, we should have re-done our analysis 
removing Star 12 from the comparison magnitude; however, in 
practice, this level of contamination is so small that we judged it was 
not worth the effort. Star 17  is obviously a non-periodic, non-variable star.

Finally, we mention the case of Star 49, which has two light curves 
shown in Figure 12. Its periodogram is also shown in Figure 9 and 
there are clearly two peaks of nearly equal height, with the shorter 
period (1.19 days) peak containing slightly more power than the 
longer period (6.27) one. These peaks are related to each other by 1/1.19 
= 1/1 - 1/6.27, showing that one is a beat frequency of the other. In 
this case we have chosen the lower power peak as the likely true 
period for a couple of reasons. First, the light curve at P=6.27 days 
looks slightly better to our eyes, even though the power in the 
periodogram is less. This happens because the power in a 
periodogram is a measure of how closely the light curve mimics a 
sine wave. Spotted variables typically do not have precisely sine 
wave-like light curves, and a slighly more scattered version of the 
actual light curve, as will appear at the beat frequecy, sometimes 
approximates a sine wave better than the less scattered, but non-sinusoidal original. We suspect that this is the case for Star 49. If not, 
this star would have the shortest period in our sample, would be one 
of the fastest rotating PMS stars known at this age, and would have a 
period uncomfortably close to one day. Another year of monitoring 
should help to resolve things for this star; here we take 6.27 days to 
be its likely rotation period.

\subsection{Some Properties of the Periodic Stars}

Perhaps the most striking fact about the stars which we have found to 
be periodic is illustrated in Figure 6. Seventeen of the 19 have 
spectral information and of these, all 17 are WTTS. It is not 
surprising that no periods are found for absorption-line stars since 
they are, as a group, not even variables, let alone periodic. Their 
absence from our periodic sample does provide some validation of 
our techniques, however, since all stars were searched for periodicity 
and the scatter in the photometry among the fainter absorption-line 
stars is certainly sufficient to generate false alarms. Somewhat more 
surprising is the total absence of CTTS's from our periodic 
sample, even though these stars are every bit as bright and variable as 
their WTTS counterparts. None of the 16 stars with known CTTS 
spectral features in our sample were found to be periodic variables.  
This is consistent with and extends the result for WTTS and CTTS in 
associations reported by Herbst \& Shevchenko (1998) who found 
that CTTS were less likely to exhibit periodic variations than WTTS 
in their large photometric data base.

This result supports the general picture of variability of PMS stars 
which has evolved over the years and was outlined in the 
introductory paragraphs of this paper. According to the canonical 
view, the WTTS vary only as a result of surface magnetic activity 
whereas the CTTS have an additional component driven by 
accretion. Evidently, on about 40\% of the WTTS in IC348 during 
our observing period, the surface inhomogeneities were sufficiently 
stable that they lasted for several months (five to fifty stellar rotations)
allowing us to detect the rotation period. Since bright spots, when 
they are seen on well studied WTTS such as V410 Tau, are both rare 
and short-lived (e.g. Vrba, Herbst \& Booth 1988) it is reasonable to 
suppose that the bulk of the variability observed, if not all of it, 
derives from large cool spots, probably at high latitudes and 
presumably associated with a strong (nearly dipole?) surface 
magnetic field. Changes in this spot pattern on timescales of weeks to 
months probably limit detection of periodicity in some of the stars. 
Others may not be oriented auspiciously for producing asymmetries 
in the light emitted from the visible hemisphere.  Active accretion on 
CTTS adds bright spots or zones to this mix which come and go on 
timescales of hours or days and mask any underlying rotational 
signature of a stable spot pattern which may be present. 
 
As in the case of the ONC, the stars found to be periodic here do not 
seem limited to or concentrated within any particular region of the 
cluster HR diagram, other than that they are TTS. Masses may be 
derived for the periodic stars from their location on the HR diagram 
and they span the range of cluster PMS masses sampled by our 
photometry (0.3 to 1.1 solar masses). There does seem to be some 
correlation between mass and rotation period in our sample, as 
shown in Figure 13. The three most massive periodic stars also have 
the shortest periods. Unfortunately, the sample is not large enough to 
allow us to conclude much from this fact at present. Discovery of a 
single more slowly rotating star with a mass greater than 0.8 solar 
masses would eliminate the "trend''. 

The frequency distribution of rotation periods in IC 348 is shown in 
Figure 14 and compared with the ONC stars more massive than 0.25 
M$_{\odot}$. Obviously the two are quite similar and no statistical test 
is needed to verify that they could be drawn from the same parent 
population. The most significant similarities, in our view, are the 
ranges (roughly a factor of 10 from $\sim$2 to somewhat less than 20 
days), the peak around 7-8 days, and the declining "tail'' of slow 
rotators beyond the peak. Also notable is the gap in the period 
distribution at about 4 days and the second peak at short period (2-3 
days). These features are obviously present in both samples but only 
verifiable as statistically significant in the larger (ONC) sample 
(Herbst et al. 2000). A small difference may be the apparent 
deficiency of extremely rapidly rotating stars in IC 348. Scaling from 
the ONC we would have expected to have found at least a couple of 
stars with P $<$ 2 days, but we found none. Of course, if our 
interpretation of the true period of star 49 is in error, then this point is 
obviated, so we cannot make too much of it at present. The principle 
result is a remarkable overall similarity between the period 
distributions in IC 348 and in the ONC. We note in passing that 
another young cluster studied at VVO with these techniques, NGC 
2264, has a substantially different period distribution (cf. Kearns \& 
Herbst 1998), possibly on account of a different age. If IC 348 and 
the ONC are both close to 1 My, then it is perhaps not surpirsing 
that their members have a similar distribution of rotation period. If IC 
348 is actually 3-4 times older than the ONC (i.e. if the closer, 
Hipparcos distance is correct) then the similarity in their rotation 
distributions is harder to understand.  

\section{Variability and IR Data}

Near infrared photometry of IC 348 was obtained in the JHK bands 
by Lada \& Lada (1995) and the original data kindly made available 
to us by the authors. We were able to match, by position, 143 of our 
151 objects with sources in their catalogue. The average difference in 
position was less than 1 arc-sec and the greatest difference was $\sim$2 
arc sec, so we are quite certain of the matches except, perhaps, in the 
cases of some stars from Table 3 with close companions. The infrared data are given for our stars in Table 2. In Figure 
15 we show the J-H vs. H-K color-color diagram based on the Lada 
\& Lada data for the stars in our sample with spectroscopic data. The 
usual boundaries, defined by limiting reddening lines are shown. 
Stars within these boundaries can be understood as having normal 
colors and normal interstellar reddening, although they may also 
have at least some contribution to their colors from a disk. Stars to 
the right of the enclosed area have excess emission in H-K which is 
generally taken to be indicative of a disk, although other sources 
(such as an infrared companion) are possible. It is interesting that all 
but one of the stars in the infrared excess region are CTTS according 
to their spectra and that, as a group, the CTTS lie further to the right 
in this diagram than do the WTTS. This is generally consistent with 
the canonical view that CTTS have disks and WTTS do not.  We 
also note that the periodic stars tend to be in the middle or towards 
the left of the enclosed area, even more so than the non-periodic 
WTTS. Perhaps this indicates that there is a class of WTTS with 
residual disks and accretion which causes some irregular variability, 
but this speculation is not supported by further analysis below.

Since spectral data are available for many of our stars, it is possible to 
go beyond the simple two-color diagram in analyzing the infrared 
data and to derive "true'' color excesses (i.e. corrected for 
extinction).  Following Hillenbrand et al. (1998) we use
$$\Delta (I-K) = (I-K) - (I-K)_o - 0.5 A_v$$
as a relatively sensitive indicator of excess emission which can be 
attributed to a disk. The observed color is computed from our I 
magnitude (Table 2) and the K magnitude of Lada \& Lada (1995). 
The intrinsic color is obtained from the spectral type given by H98 or 
Luhman et al. (1998) using the calibration of Kenyon \& Hartmann (1995). 
The visual extinction is computed from the color excess in V-I, again 
using the spectral-color relation of Kenyon \& Hartmann and a standard 
extinction law. The excess I-K emission is plotted as a function of 
$\sigma_{var}$ in Figure 16. Seven of the 8 stars with $\Delta (I-K) 
> 0.4$ are CTTS according to their spectra. So, in spite of the scatter 
introduced by variability, errors in photometry and spectral type, and 
unknown influences such as inner disk holes, disk inclination and unresolved cool
companions (see Hillenbrand et al. 1998 for a discussion of possible errors) 
we find that this indicator is reasonably good at picking out stars 
with disks (as indicated by their H$\alpha$ equivalent widths). 

Figure 16 also shows that there is almost no correlation between 
variability and IR excess emission (at least in I-K) in our data, 
although it is true that the WTTS as a group are less variable and 
have little or no excess while the CTTS as a group are more variable 
and tend to have IR excesses. Probably the errors and, in particular, 
the variability obscure whatever correlation might exist. It is 
interesting that the periodic stars cluster around $\Delta (I-K)  = 0$ 
with one exception, star 73, the periodic star with, by far, the 
largest amplitude. Two explanations for this star's somewhat 
anomalous location on the diagram suggest themselves. This could 
be a star with a hot spot, which would explain the large photometric 
amplitude, rather poorly defined light curve, and IR excess emission, 
although that would imply that at least this one WTTS does have an 
accretion disk surrounding it. Alternatively, this may be a star with a 
very large cool spot and favorable orientation to produce such a large 
amplitude. In that case, the IR excess might not arise from a disk at 
all, but from the large surface area of the cool spot, which will radiate 
substantially in K. A simple quantitative analysis 
indicates that the size of the excess could be explained by this 
phenomenon within the errors of its determination. We note that 
there is, in fact, a weak correlation among the stars with periods, 
between amplitude of variation and IR excess, which could also be 
explained by increasing amounts of cool spot emission in K with 
increasing spot area. We have no explanation for the fact that the 
non-periodic WTTS tend to cluster around negative values of  
$\Delta (I-K)$, but this does not support the speculation based on 
their location in the J-H versus H-K diagram that they may have 
some residual disk emission and accretion.

Finally, we discuss the relationship between excess IR 
emission and rotation period. Edwards et al. (1993) found a good 
correlation between period and H-K 
excess ($\Delta (H-K) $) for a sample of stars in Taurus and Orion and took this as 
evidence of a role for disks in influencing stellar rotation. Herbst et al. 
(2000) showed that there was a weak, but significant, correlation 
between both $\Delta (H-K) $ and $\Delta (I-K )$ in the ONC, based on a much expanded 
data set. However, it is true that as data have accumulated, the 
original correlation of Edwards et al. (1993) has become less well 
defined. In IC 348 there is essentially no 
correlation between IR excess emission and rotation period; in fact,
since all of the periodic stars are WTTS, it is not 
surprising that none of them (with the possible exception of star 73) 
show evidence for disk emission. Although no correlation is present,
 the situation is not that  dissimilar to what Herbst et al. 
(2000) found in the ONC or what the Edwards et al.  (1993) distribution 
looks like. In particular, both of these studies found many stars at all 
rotation periods with no evidence for IR excess emission. The 
difference is that in the ONC there are some stars with IR excesses 
and they all have longer periods than about 7 days. If star 73 in IC 348 is 
actually an accreting star, then its period of 7.47 days is consistent 
with the previous result that accreting stars have periods in the longer 
period peak of the bimodal distribution. 

This study reveals a potentially important factor concerning the possible connection 
between rotation and disks which has not previously been 
recognized. Namely, it is evidently easier to find rotation periods for 
non-accreting stars than for those with accretion disks. Therefore, any 
correlation between accretion and rotation will be harder to establish 
than would otherwise be the case. Probably what is needed, in fact, is 
to resort to v sin i measurements so that a complete sample of WTTS 
and CTTS can be compared. Given enough stars to average the 
effects of inclination, this test should show definitively whether stars 
with accretion disks really do tend to rotate more slowly than stars 
without them as one might expect under the disk-locking hypothesis 
(cf. K\"onigl 1991; Ostriker \& Shu 1995). A v sin i study in the ONC is 
nearing completion (Rhode et al. 2000) and one in IC 348 would be 
a valuable undertaking.

\section {Final Remark and Some Questions Raised}

We have discovered about 50 new variable stars in the extremely 
young cluster IC 348 and studied their variations over a period of 
several months.  All of the variables for which we have spectral 
information are either CTTS or WTTS. It is, perhaps, surprising that 
none of the G-type absorption stars turned out to be definite variables 
(star 13 is possibly variable), since 
there are examples of such objects in associations (e.g. SU Aur). As 
a group, the CTTS are more highly variable than the WTTS, have 
significant IR excess emission, and show irregular variations, as 
opposed to periodic ones. All of this is consistent with the canonical 
view that the critical difference between CTTS and WTTS is an 
accretion disk. Since CTTS are definitely present in this young 
cluster, with an age of 1 - 3 million years, this demonstrates that 
some accretion disks can last that long. However, since WTTS are at 
least as numerous as CTTS in the cluster (H98) and, in fact, 
outnumber the CTTS by 3 to 1 in our sample, it is also clear that the 
majority of stars in the central portion of IC 348 with masses less 
than 1 M$_\odot$ have lost their disks in  1 - 3 million years.  
The timescale for disks to disappear around stars 
a bit more massive than 1 M$_\odot$ must be $<$ 0.3 Myr since 
none of the G-type stars have any evidence of an accretion disk. 
These timescales are somewhat shorter than what is often assumed 
for solar mass stars when processes such as planet building, for 
example, are studied (c.f. Boss 2000).

While this study is generally consistent with the canonical picture of
PMS variability, many questions remain. Among these is the issue of the
non-periodic WTTS. Are these systems with a small amount of accretion, or
are they stars with changing spot patterns during the observational epoch?
Additional years of photometric monitoring may answer that question. What is the rotation
velocity of the CTTS in this cluster and do they, as a group, rotate slower
than the WTTS? A spectroscopic v sin i study is required to address this issue.
What is the long term (years) nature of the CTTS and the WTTS variables? 
Further study will reveal how the spot patterns change or remain stable with
time and whether any FUors, EXors or UXors are to be found in this cluster. We
intend to keep monitoring the cluster at Van Vleck Observatory for a few more
seasons, at least, to address these issues.

\paragraph{Acknowledgments.}

It is a pleasure to thank the W. M. Keck Foundation for their support of the Keck Northeast Astronomy Consortium, which provided funding for some of
the equipment used in this work. The new CCD was made possible through a 
grant from the NSF which we gratefully acknowledge. We thank G. Herbig for making a FORTRAN program available for this work and for helpful discussions. We thank C. and E. Lada for making their infrared photometry available to us and we thank the many student observers at Wesleyan who helped obtain the data.
NASA has supported this research through a grant from its "Origins of Solar Systems'' program. One of us (W.H.) was on sabbatical leave at the Max-Planck-Institut f\"ur Astronomie in Heidelberg, Germany, during the time of writing this paper and wishes to thank his sponsors I. Appenzeller and R. Mundt and the Institute staff for their support and hospitality. 

\newpage 
\subsubsection*{References}

\noindent Adams, N. R., Walter, F. J. \& Wolk, S. J. 1998, \aj 116, 237\\
Attridge, J. M. \& Herbst, W. 1992, \apj 398, L61\\
Bertout, C. 1989, \araa 27, 351\\
Bessell, M. S. 1990, \pasp 102, 1181\\
Boss, A. 2000, in Proceedings of the European Conference on Disks, Planetesimals and Planets held in Tenerife, Spain, Jan. 2000 (ASP, San Fransisco), edited by F. Garz\'on, C. Eiroa, D. de Winter \& T. J. Mahoney, in preparation. 
Bouvier, J. et al. 1999, \aap 349, 619\\
Choi, P. I. \& Herbst, W. 1996, \aj 111, 283\\
D'Antona, F. \& Mazzitelli, I. 1994, \apjs 90, 467\\
Duchene, G., Bouvier, J., \& Simon, T. 1999,  \aap 343, 831\\
Eaton, N. L., Herbst, W. \& Hillenbrand, L. A. 1995, \aj 110, 1735\\
Edwards et al. 1993, \aj 106, 372\\
Grinin, V. P. 1994,in "The Nature and Evolutionary Status of Herbig Ae/Be Stars'',Astronomical Society of the Pacific Conference Series, Vol 62 (ASP, San Francisco), edited by P.S. The, M. Perez and E. P. J. Van den Heuval, p. 63\\
Haro, G., Chavira, E., \& Mendoza, E. 1960, \aj 65, 490\\
Hartmann, L. 1998, "Accretion Processes in Star Formation'' (Cambridge U. Press, Cambridge U.K.
Hartmann, L. \& Kenyon, S. J. 1996, \araa 34, 207\\
Herbig, G. H. 1960, \apjs 4, 337\\
Herbig, G. H. 1977, \apj 217, 693\\
Herbig, G. H. 1998, \aj 497, 736 (H98)\\
Herbst, W. 1994, in "The Nature and Evolutionary Status of Herbig Ae/Be Stars''
Astronomical Society of the Pacific Conference Series, Vol 62 (ASP, San Francisco), edited by P.S. The, M. Perez and E. P. J. Van den Heuval, p. 35\\
Herbst, W., Herbst, D. K., Grossman, E. J. \& Weinstein, D. 1994, \aj 108, 1906\\
Herbst, W., Rhode, K. L., Hillenbrand, L. A., \& Curran, G. 2000, \aj 119, 261\\
Herbst, W. \& Shevchenko, V. S. 1998, \aj 116, 1419\\
Herbst, W. \& Wittenmyer, R. 1996, \baas 189, 4908\\
Hillenbrand, L. A. 1997, \aj 113, 1733\\
Hillenbrand, L. A., Strom, S. E., Vrba, F. J., \& Keene, J. 1992, \apj 397, 613\\
Hogg, H. S. 1972, in "Variable Stars in Globular Clusters and in Related Systems", Proceedings of IAU Colloquium 21 (D. Reidel, Dordrecht), edited by H. D. Fernie, p. 3\\
Horne, J. H. \& Baliunas, S. L. 1986, \apj   302, 757\\
Joy, A. H. 1945,\apj 102, 168\\
Kearns, K. E. \& Herbst, W. 1998, \aj 116, 261\\
Kenyon, S. J. \& Hartmann, L. 1995, \apjs 101, 117\\
K\"ognil, A. 1991, \apjl 370, L39\\ 
Lada, E.A. \& Lada, C.J. 1995, \aj 109, 1682\\
Landolt, A. U. 1992, \aj 104, 340\\
Luhman, K. L., Rieke, G.H., Lada, C.J., \& Lada, E.A. 1998, \apj 
508, 347\\
Mahdavi, A. \& Kenyon, S. J. 1998, \apj 497, 342\\
Makidon, R. B., Adams, M. T., Strom, S. E., Kearns, K., Herbst, W. \& Jones, B. F. 1996, \baas 189.4910\\
Mandel, G. N. \& Herbst, W. 1991, \apj 383,L75\\
Natta, A., Prusti, T., Neri, R., Thi, W. F., Grinin, V. P. \& Mannings, V. 1999, \aap 350, 541\\
Ostriker, E. \& Shu, F. 1995 \apj, 447, 813\\
Parenago, P. P. 1954, Trudy Gosud. Astron. Sternberga, 25, 1\\
Petrov, P. P. \& Herbig, G. H. 1992, \apj 392, 209\\
Preibisch, T., Zinnecker, H.A., \& Herbig, G.H. 1996, \aap 310, 456\\
Rhode, K. L., Herbst, W. \& Mathieu, R. D. 2000, in preparation\\
Scargle, J. D. 1982, \apj   263, 835\\
Scholz, R.-D., Brunzendorf, J., Ivanov, G., Kharchenko, N., Lasker, 
B., Meusinger, H., Preibisch, T., Schilbach, E. \& Zinnecker, H. 1999, 
\aaps 137, 305\\
Stassun, K. G., Mathieu, R. D., Mazeh, T., \& Vrba, F. J. 1999, \aj 117, 2941\\
Strom, S. E., Strom, K. M., Yost, J., Carrasco, L. \& Grasdalen, G. 1972, \apj 173, 353\\
Trullols, E. \& Jordi, C. 1997, \aap 324, 549 (TJ)\\
Vrba, F. J., Herbst, W. \& Booth, J. F. 1988, \aj 96, 1032\\
 \newpage 
\begin{figure}
\plotone{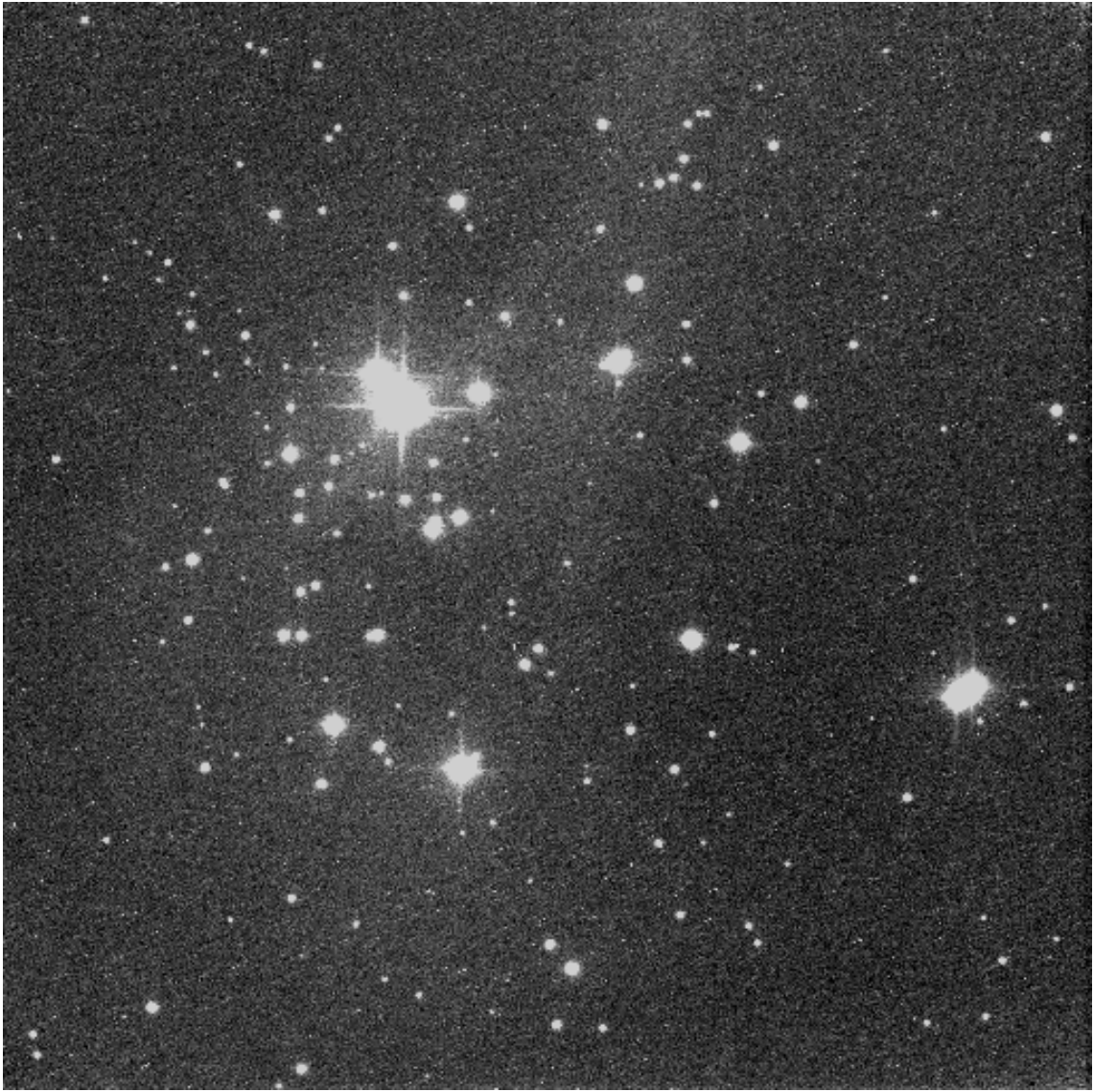} 
\figcaption{ A five minute image of IC 348 obtained with the 0.6 m 
telescope at VVO through a Cousins I filter, showing the part of the 
cluster monitored for variability. The field is 10.2$\arcmin$ on a side 
with North at the top and East to the left.}  
\end{figure}

\begin{figure} 
\epsscale{0.8}
\plotone{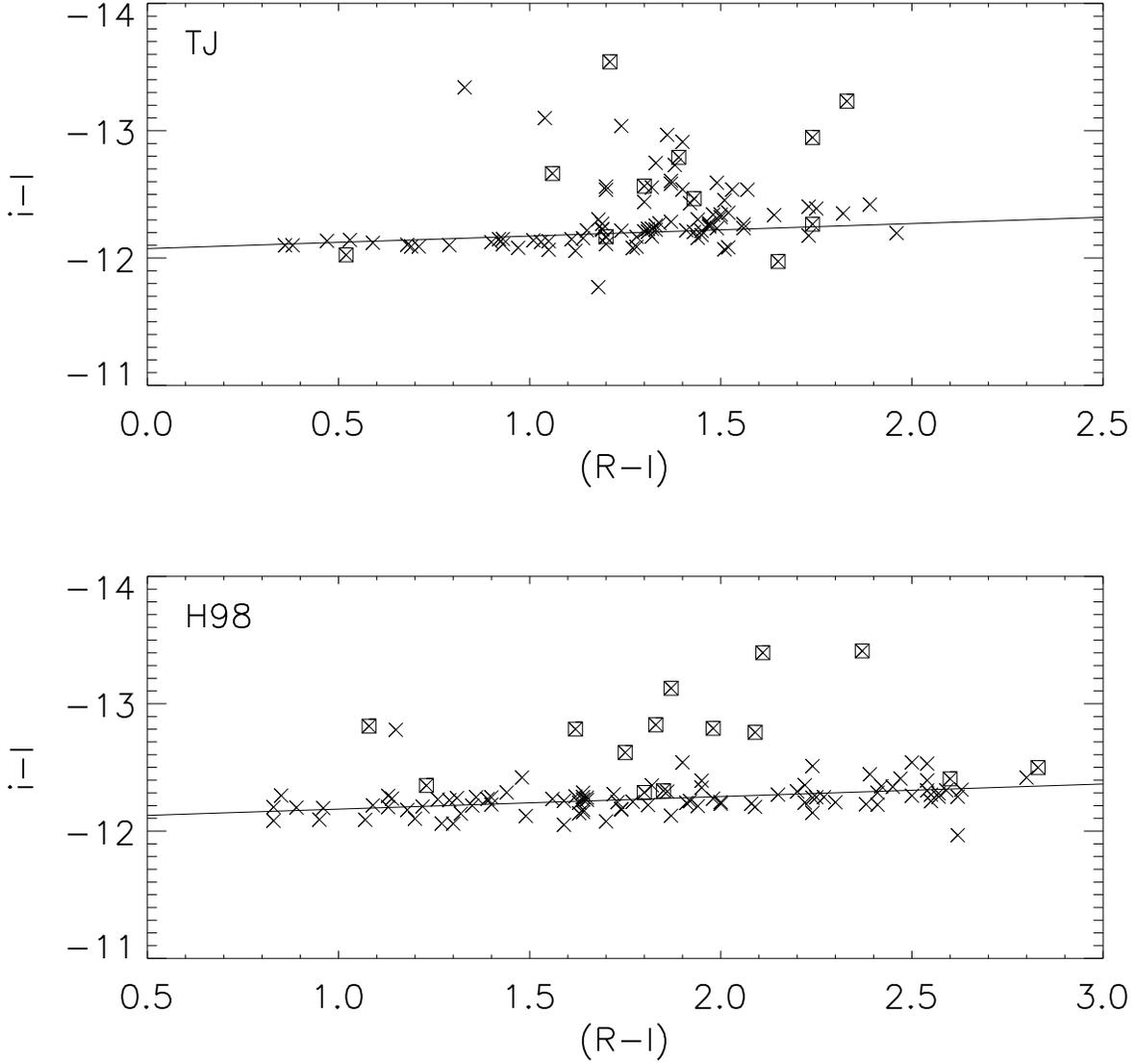} 
\vskip 2 cm
\figcaption{The difference between mean instrumental magnitude (i) as 
measured by us and standard magnitude on the Cousins system (I) 
measured by Trullols \& Jordi (top) and Herbig (bottom) as a 
function of color. Boxed points are visual binaries from Table 3 and 
may have contaminated photometry. A least squares fit to the 
uncontaminated Herbig data is shown as the line in both panels.} 
\epsscale{1.0} 
\end{figure}

\begin{figure} 
\plotone{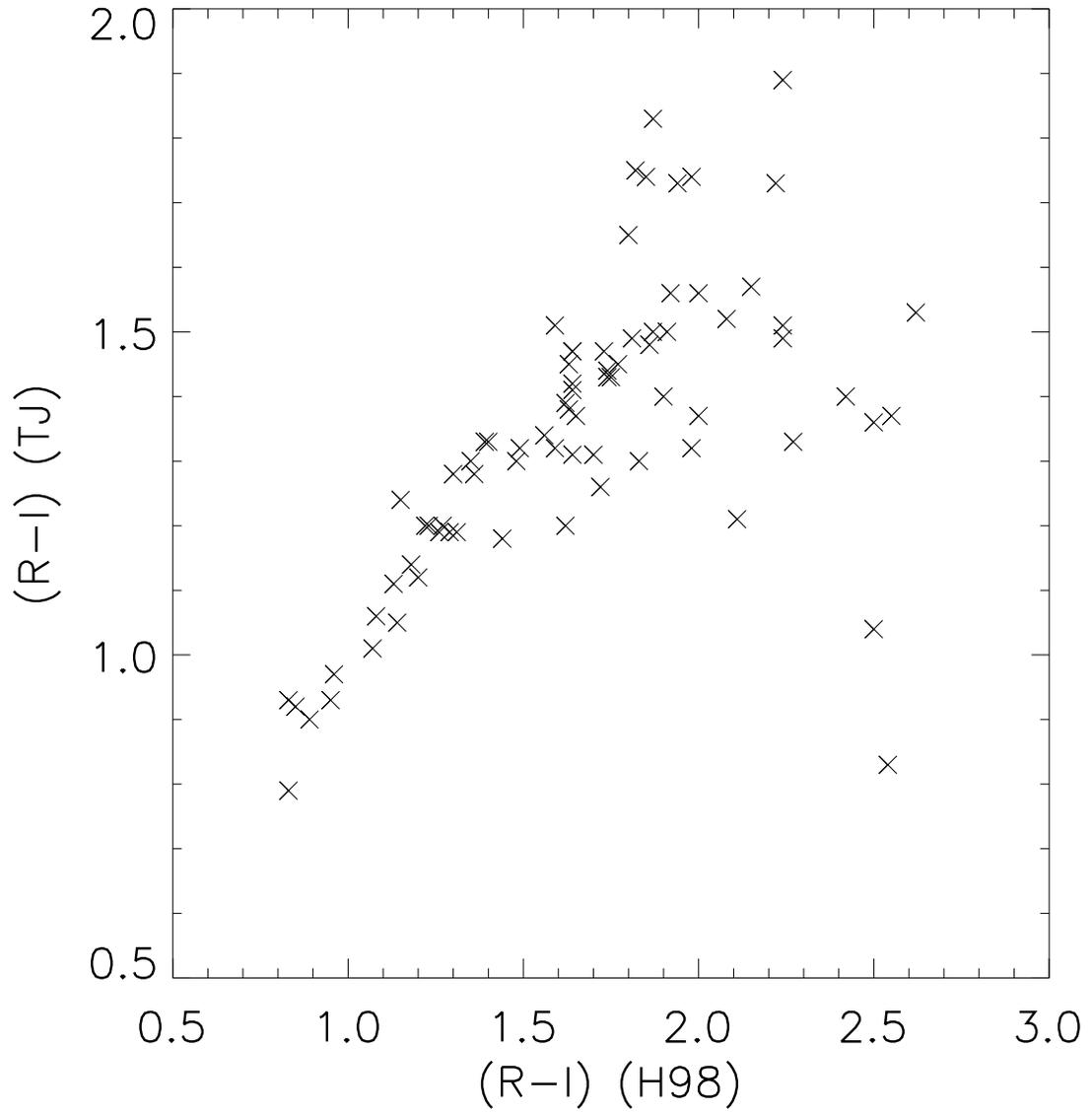} 
\figcaption{ R-I as measured by Trullols \& Jordi (TJ) plotted against R-I as 
measured by Herbig (H98) for stars which were monitored by us.}  
\end{figure}

\begin{figure} 
\plotone{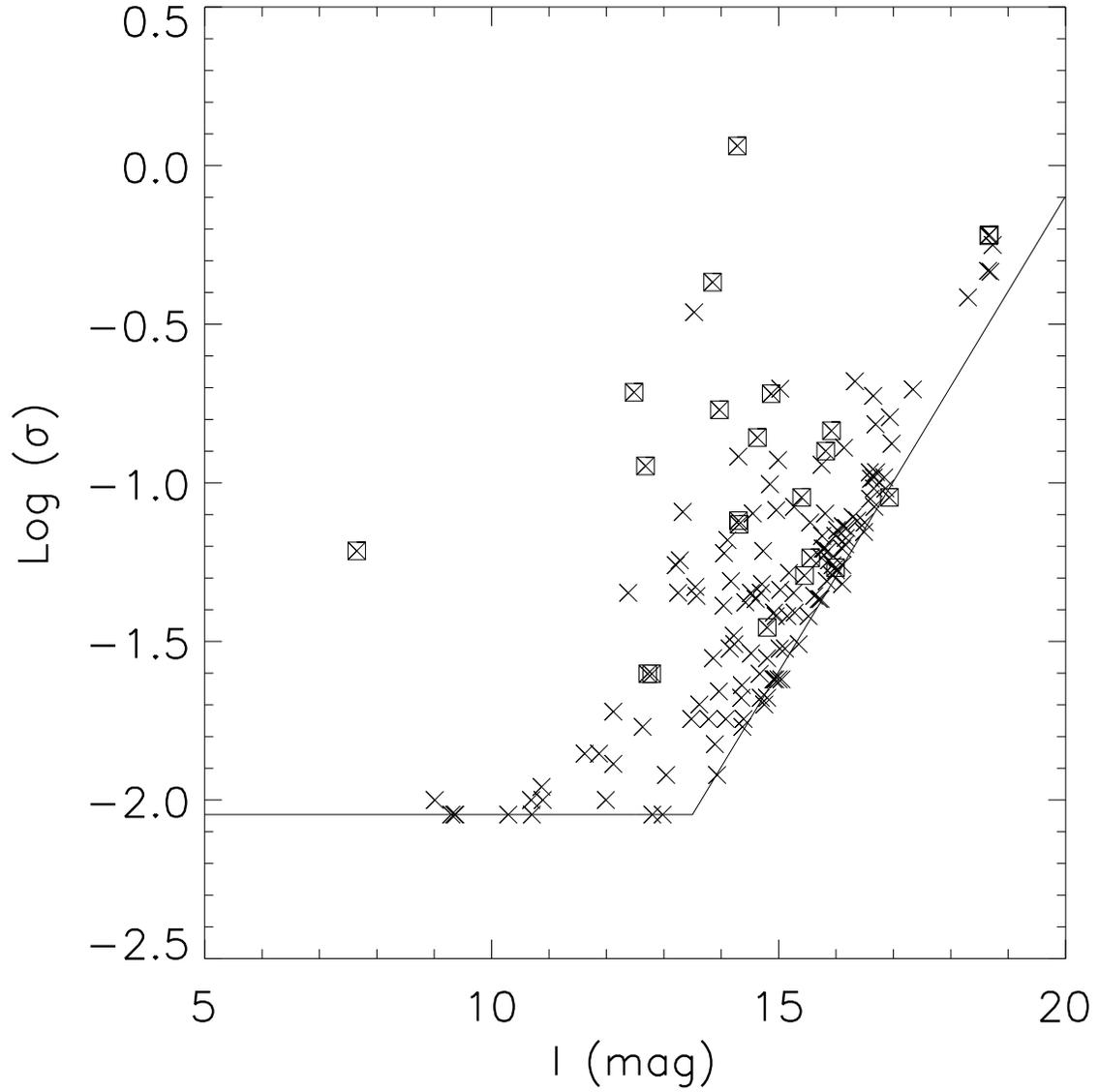} 
\figcaption{The logarithim of the scatter in the data for each star, as 
measured by its $\sigma$, as a function of average brightness. Boxed 
stars are from Table 3, which means their photometry is possibly contaminated by 
light from a companion.}  
\end{figure}

\begin{figure} 
\plotone{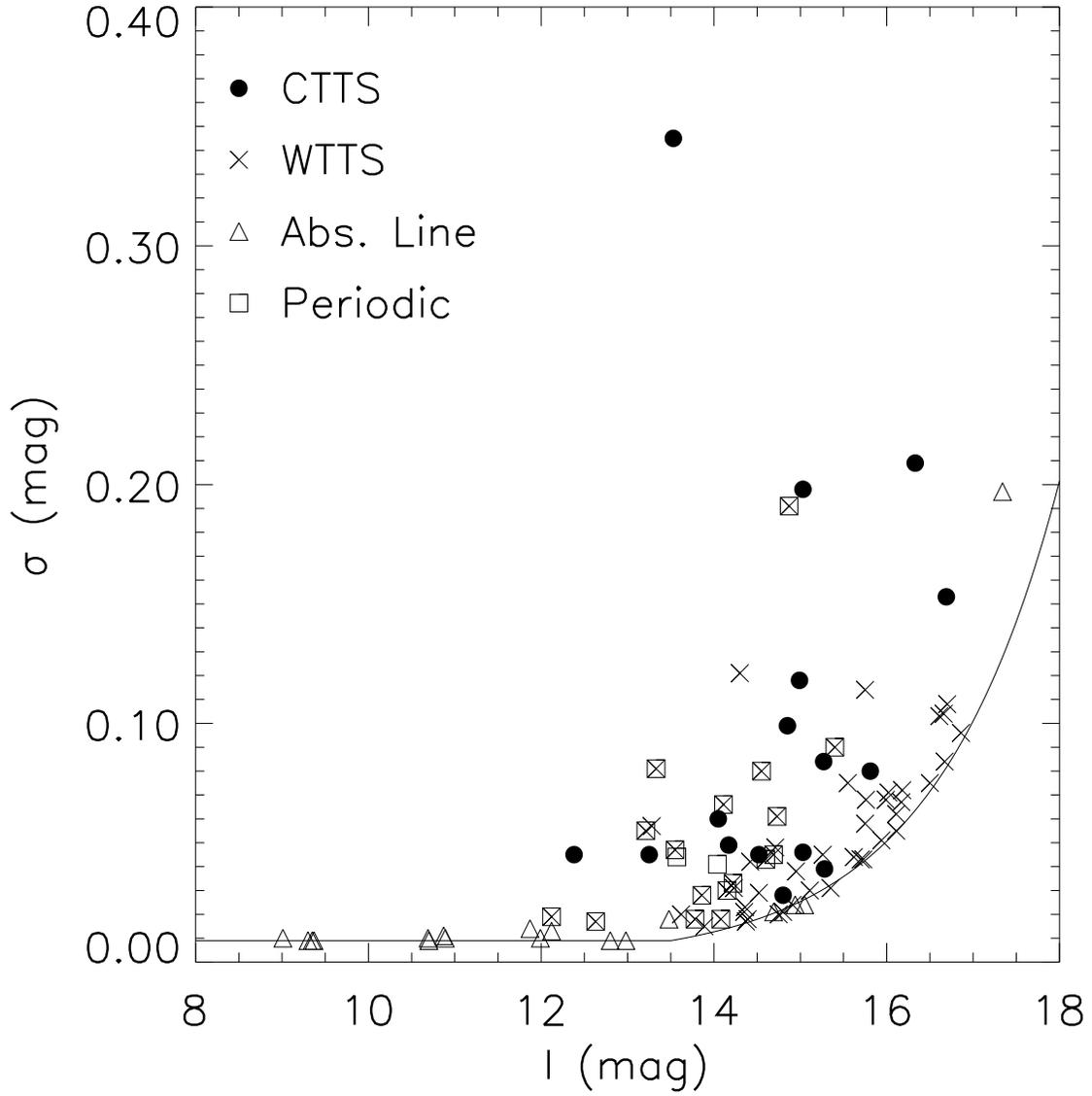} 
\figcaption{The scatter in the photometry of a star as a function of its 
brightness. Different symbols correspond to different spectral 
categories, as indicated. Boxed points are periodic variables. The line 
is the lower limit to the variability as determined from Figure 4.}  
\end{figure}

\begin{figure} 
\plotone{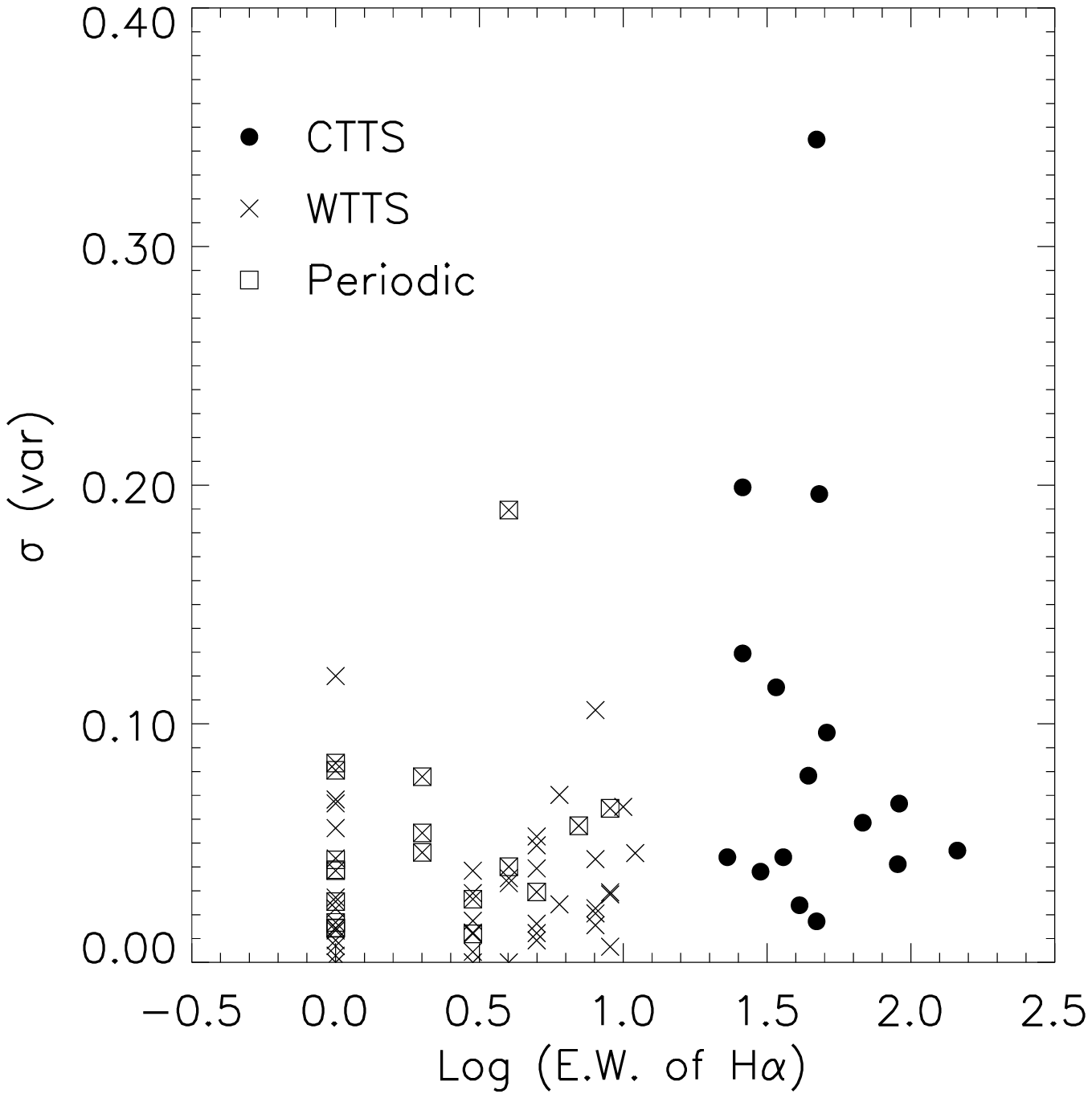} 
\figcaption{The portion of the star's scatter which can be attributed to 
actual variability is plotted against the equivalent width of the 
H$\alpha$ line.}  
\end{figure}

\begin{figure} 
\epsscale{0.8}
\plotone{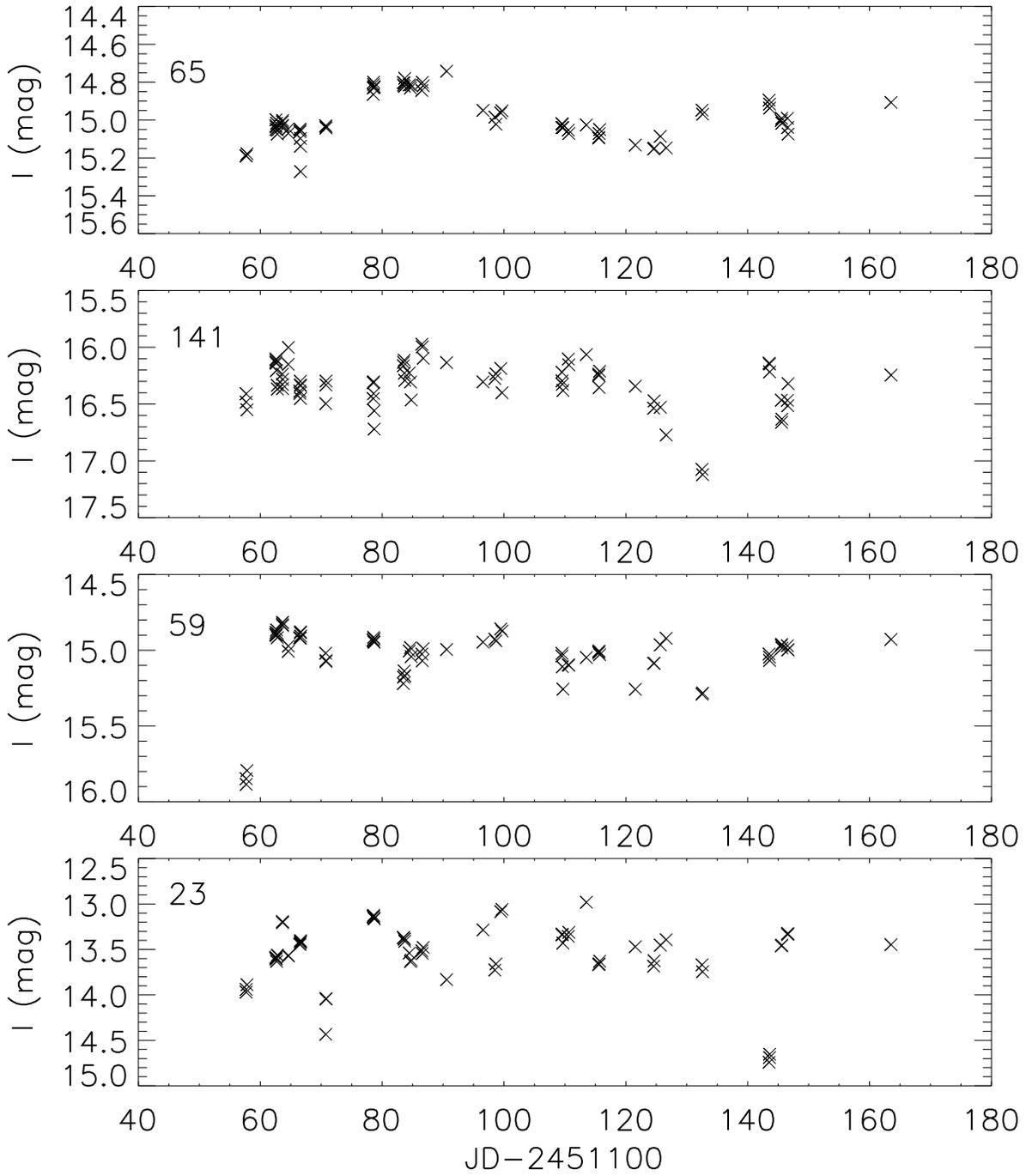} 
\vskip 2 cm
\figcaption{Light curves of four CTTS.}  
\end{figure}

\begin{figure} 
\plotone{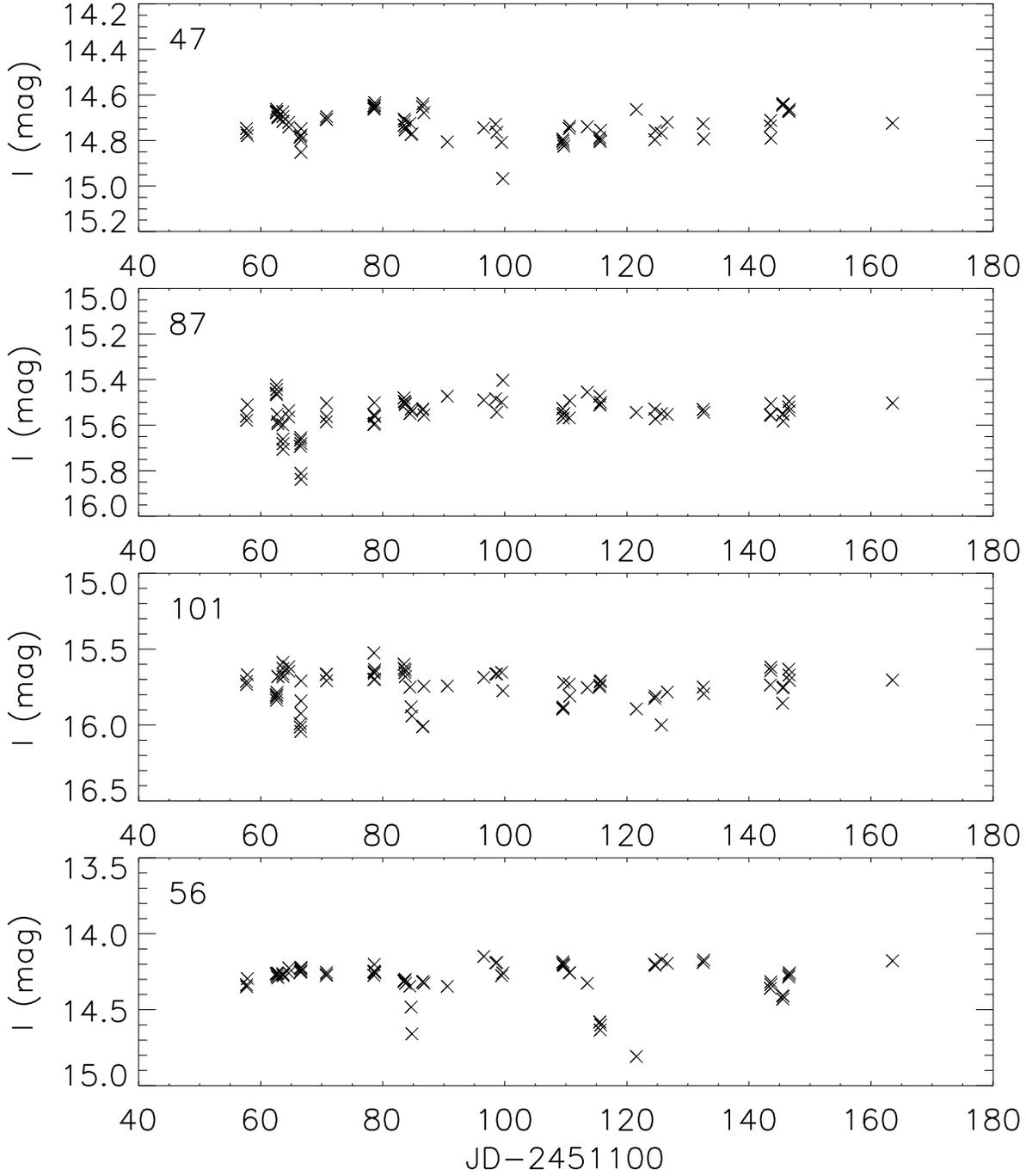} 
\vskip 2 cm
\figcaption{Light curves of four non-periodic WTTS of large 
amplitude.} 
\epsscale{1.0} 
\end{figure}

\begin{figure} 
\plotone{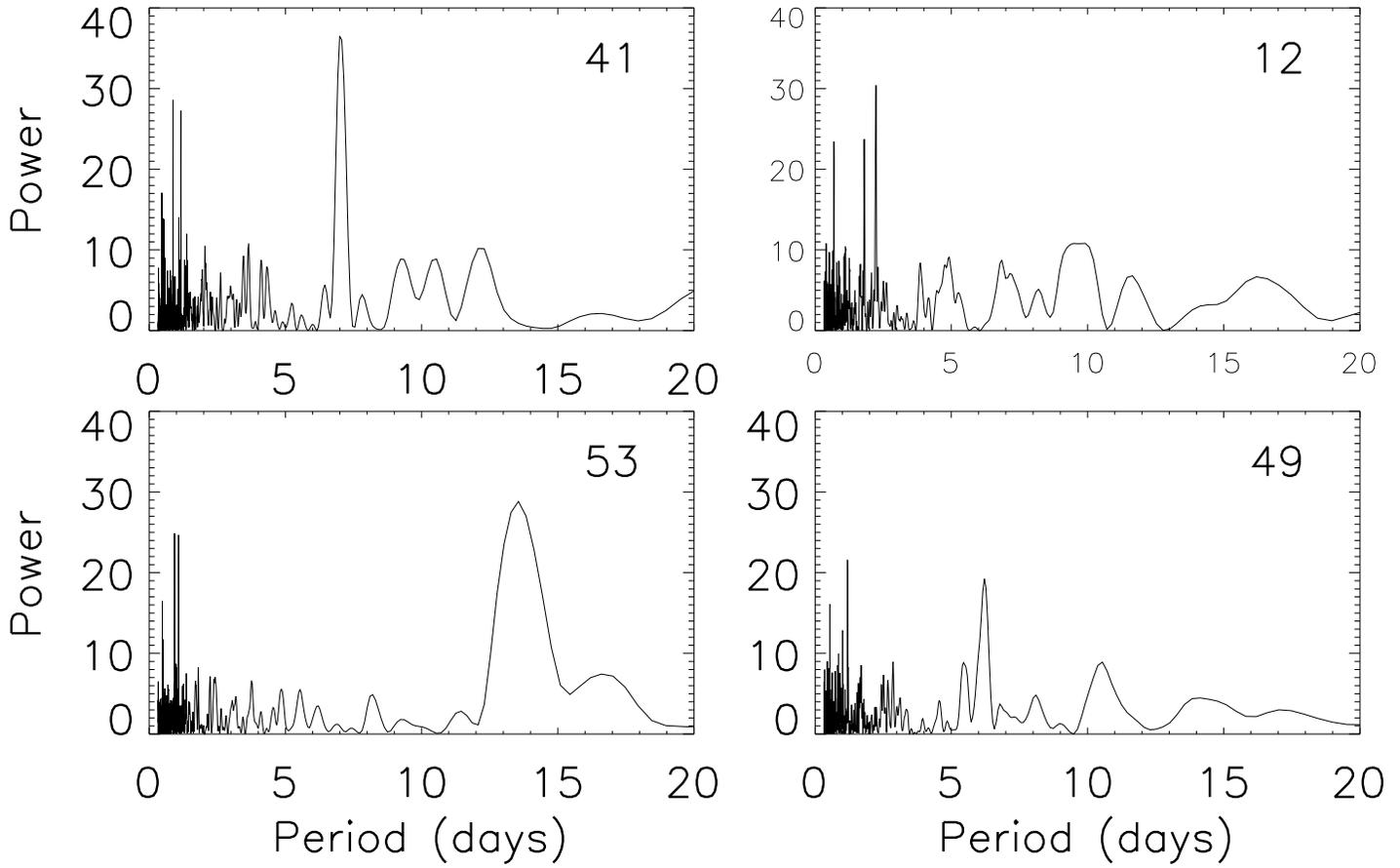} 
\figcaption{A sample of four periodograms of stars identified as 
periodic.  Stars 41 and 49 have, respectively, the largest and smallest 
peaks considered significant. Stars 12 and 53 have, respectively, the 
shortest and longest periods in our sample.}  
\end{figure}

\begin{figure} 
\epsscale{0.8}
\plotone{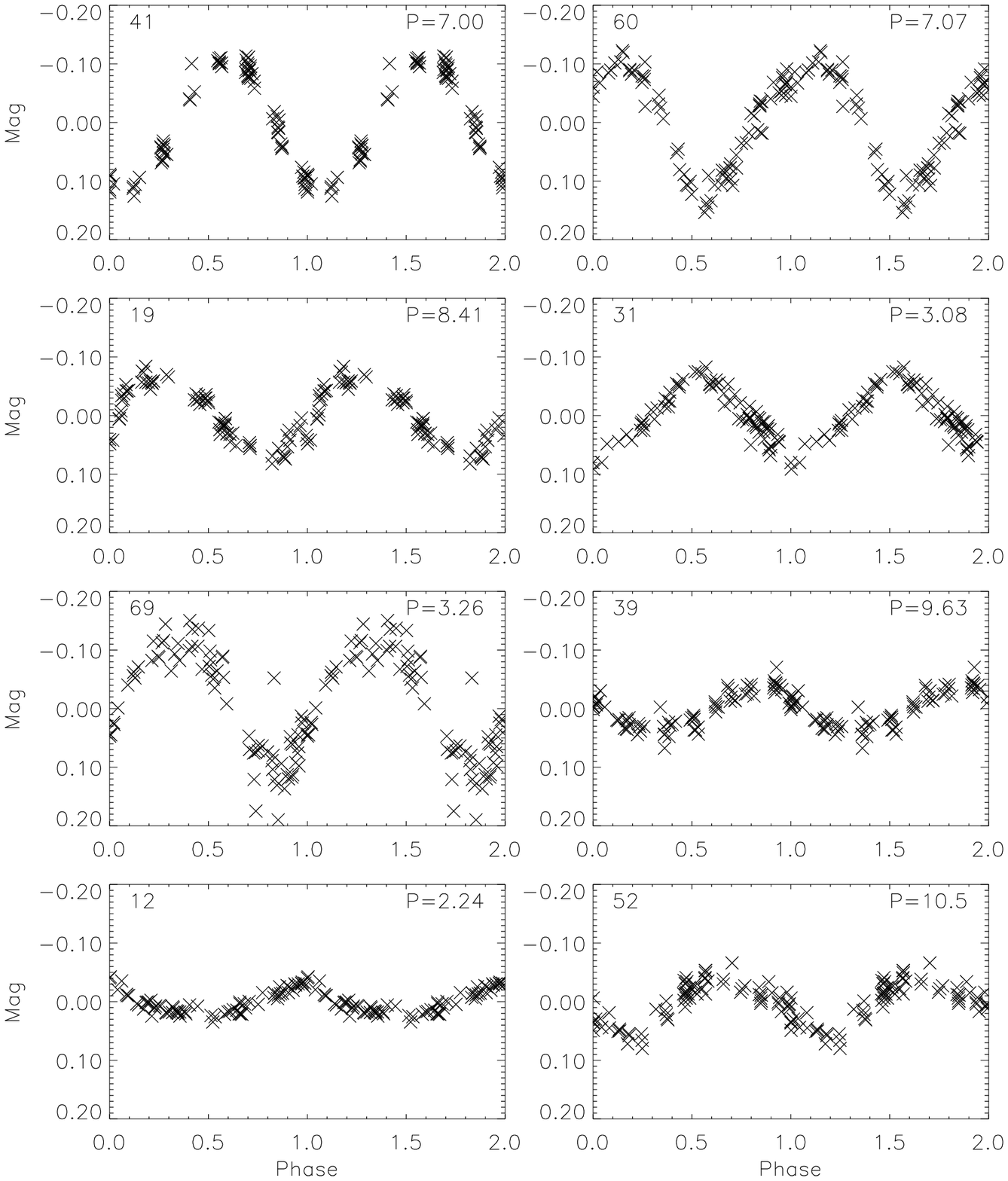} 
\vskip 2 cm
\figcaption{Light curves of periodic variables.}  
\end{figure}

\begin{figure} 
\plotone{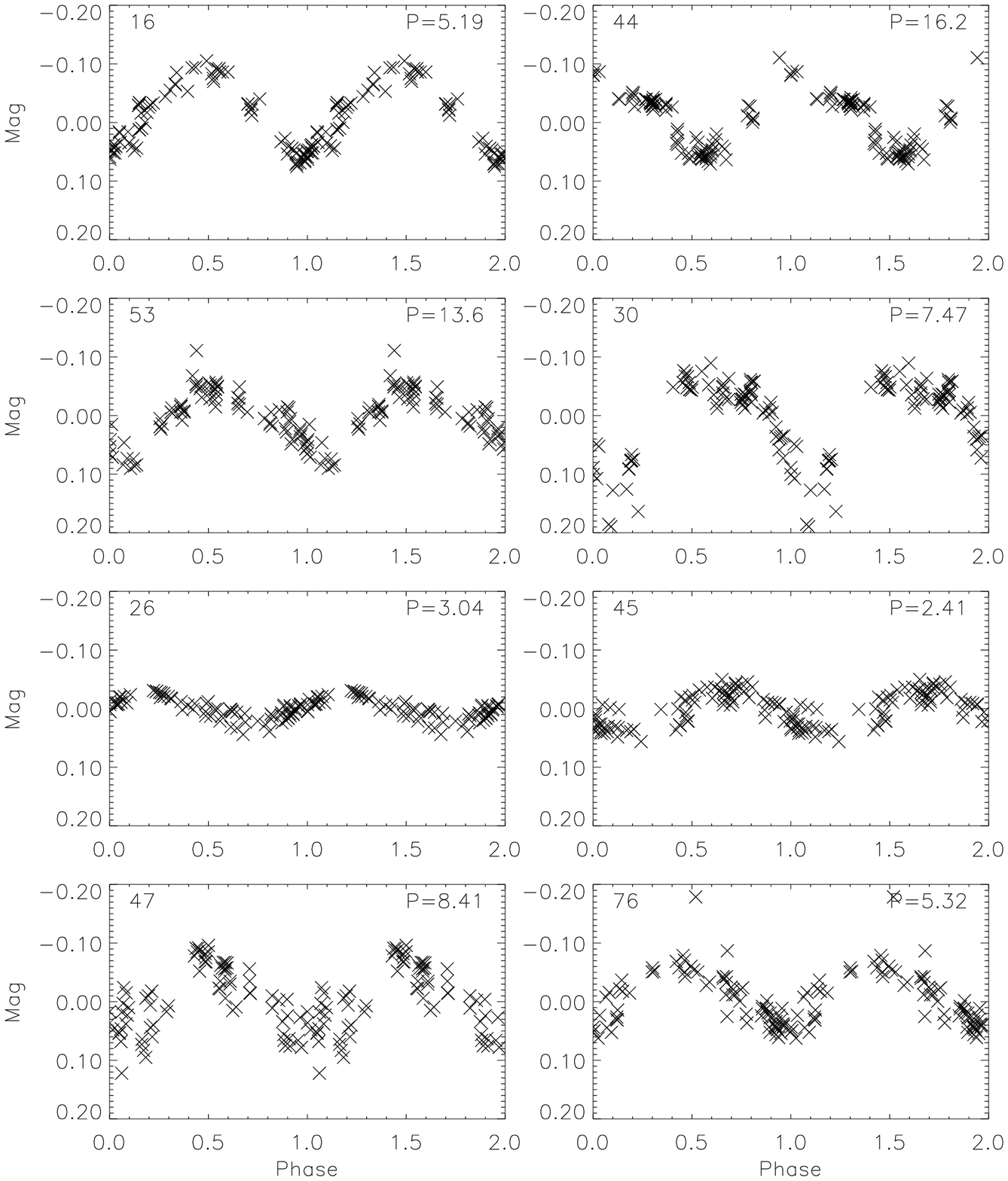} 
\vskip 2 cm
\figcaption{Additional light curves of periodic variables.} 
\epsscale{1.0} 
\end{figure}

\begin{figure} 
\plotone{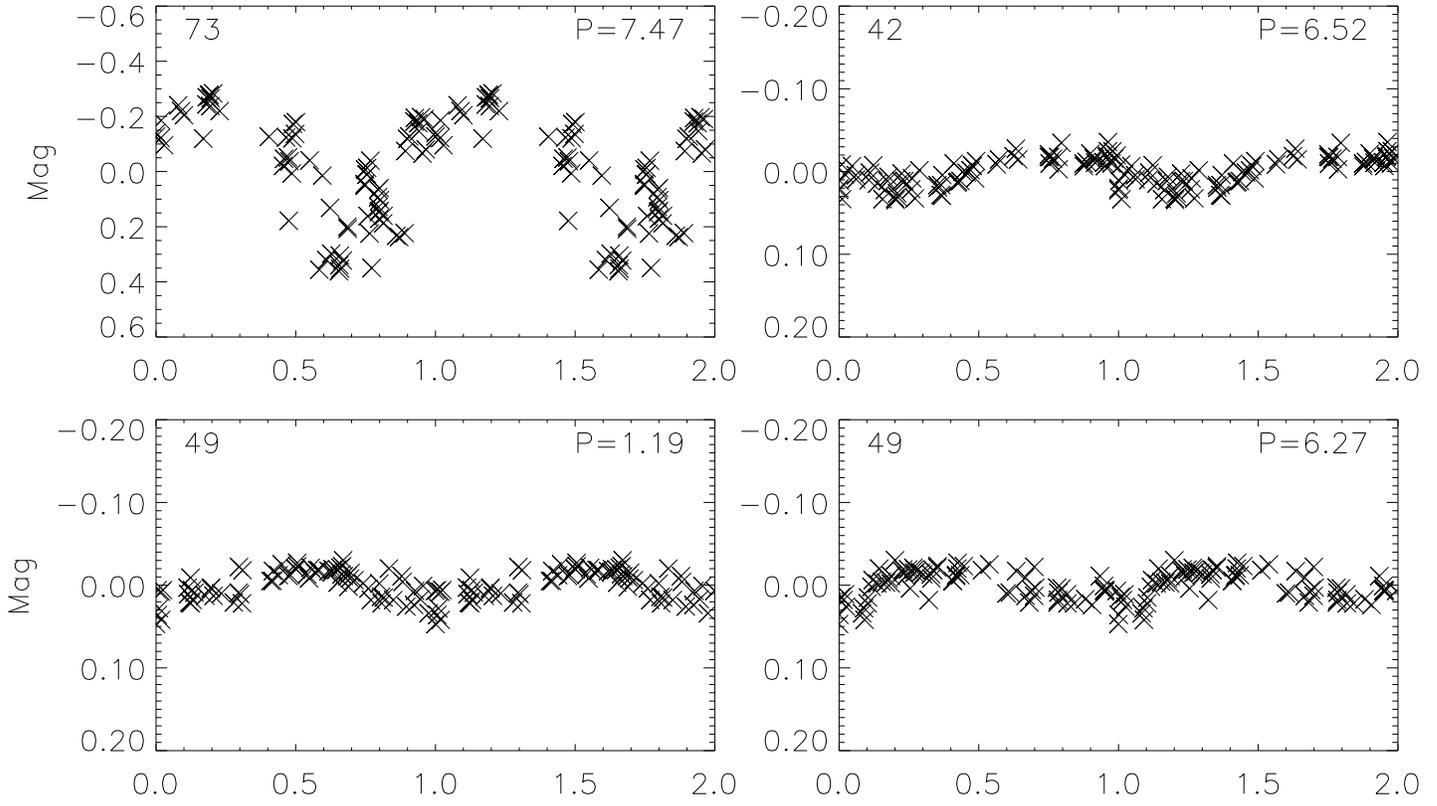} 
\figcaption{Additional light curves of periodic variables. Star 49 is 
shown at two possible periods. We adopt the longer one here.}  
\end{figure}

\begin{figure} 
\plotone{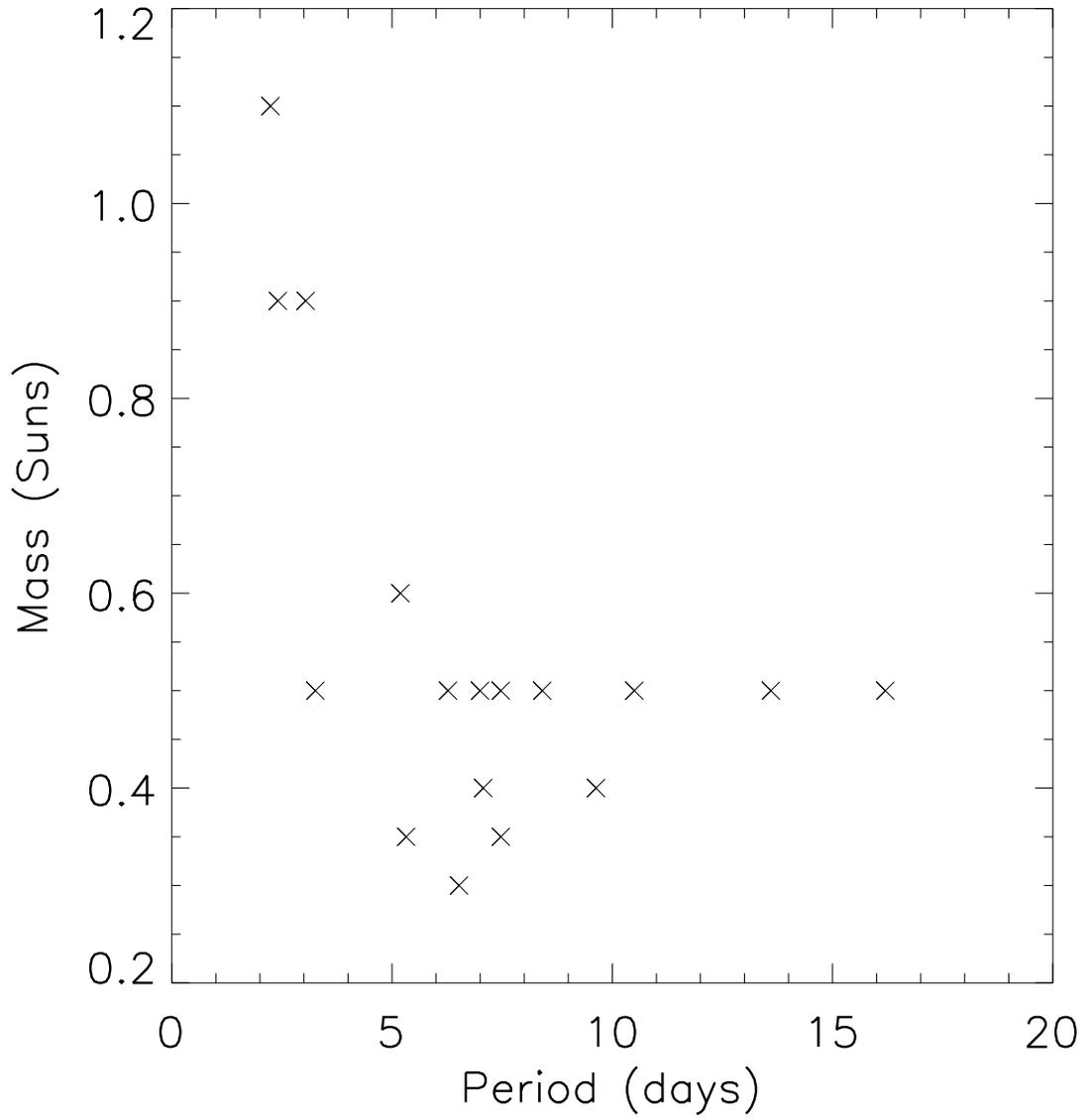} 
\figcaption{Mass versus rotation period for the periodic stars in our 
sample.}  
\end{figure}

\begin{figure} 
\plotone{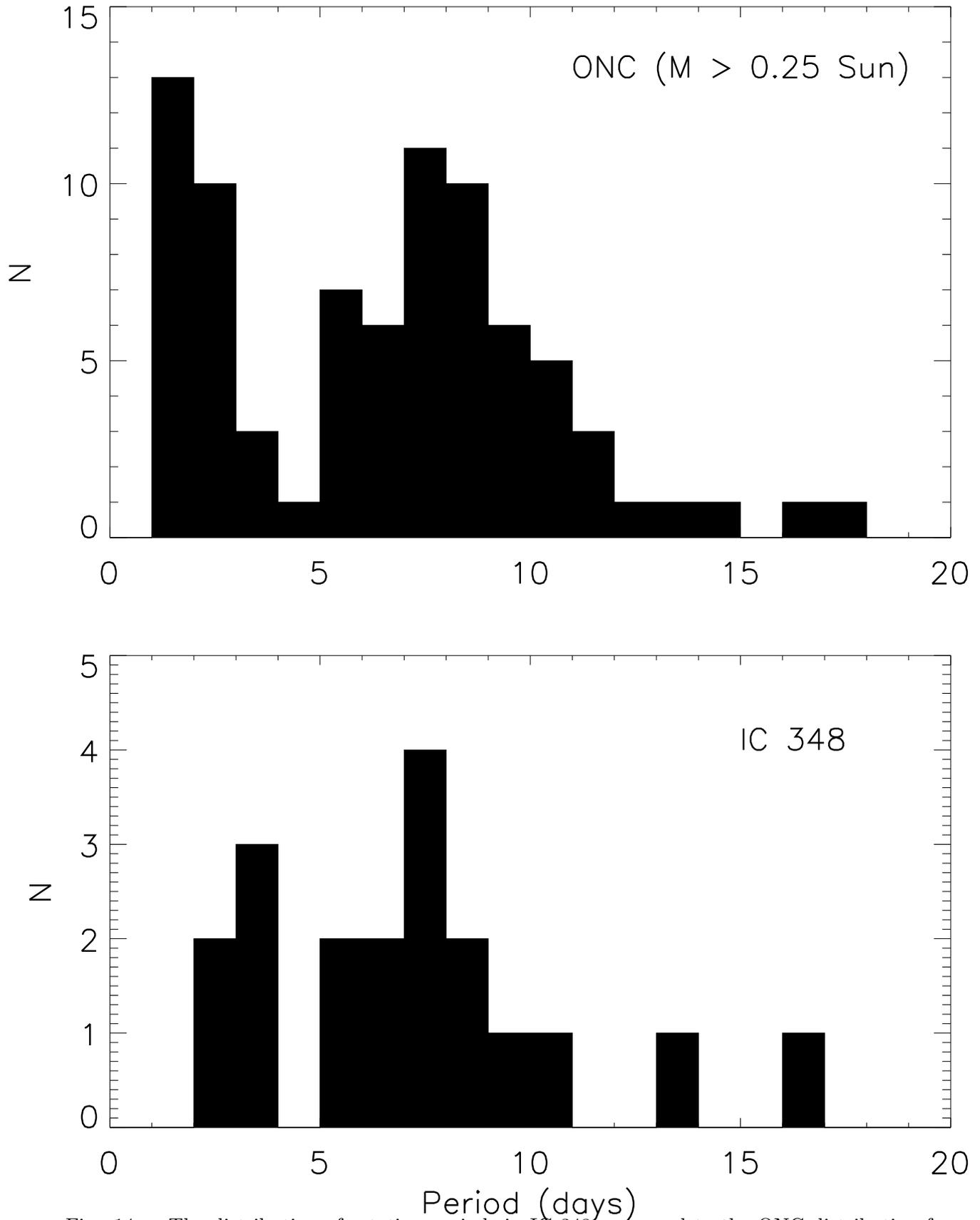} 
\figcaption{The distribution of rotation periods in IC 348 compared to 
the ONC distribution for stars more massive than 0.25 solar masses.}  
\end{figure}

\begin{figure} 
\plotone{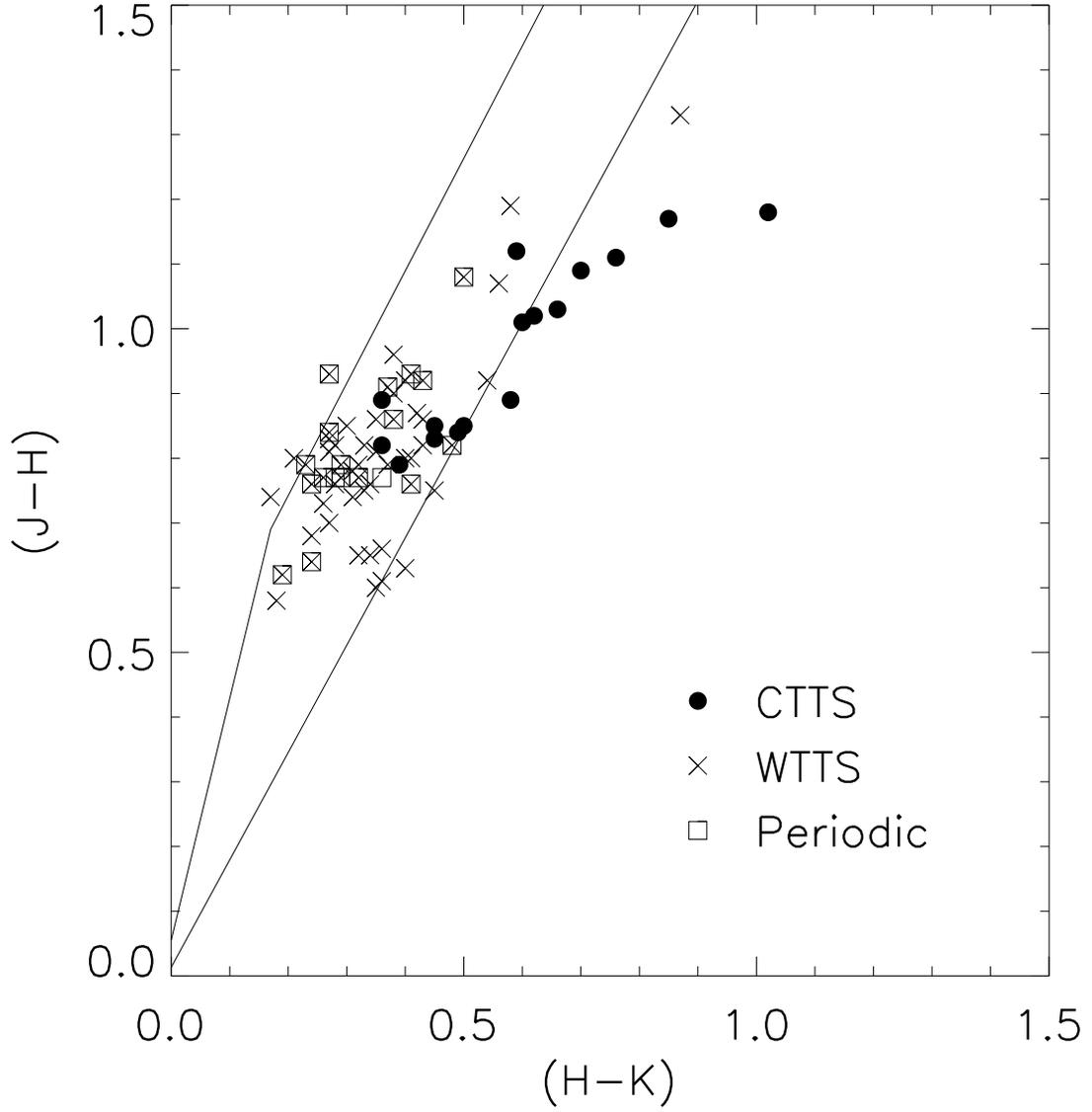} 
\figcaption{J-H vs. H-K for stars in our field, as measured by Lada \& 
Lada (1995). A portion of the color-color relation for normal stars 
and extreme reddening lines are shown. Stars to the right of the 
bounded region have colors indicative of circumstellar disks.}  
\end{figure}

\begin{figure} 
\plotone{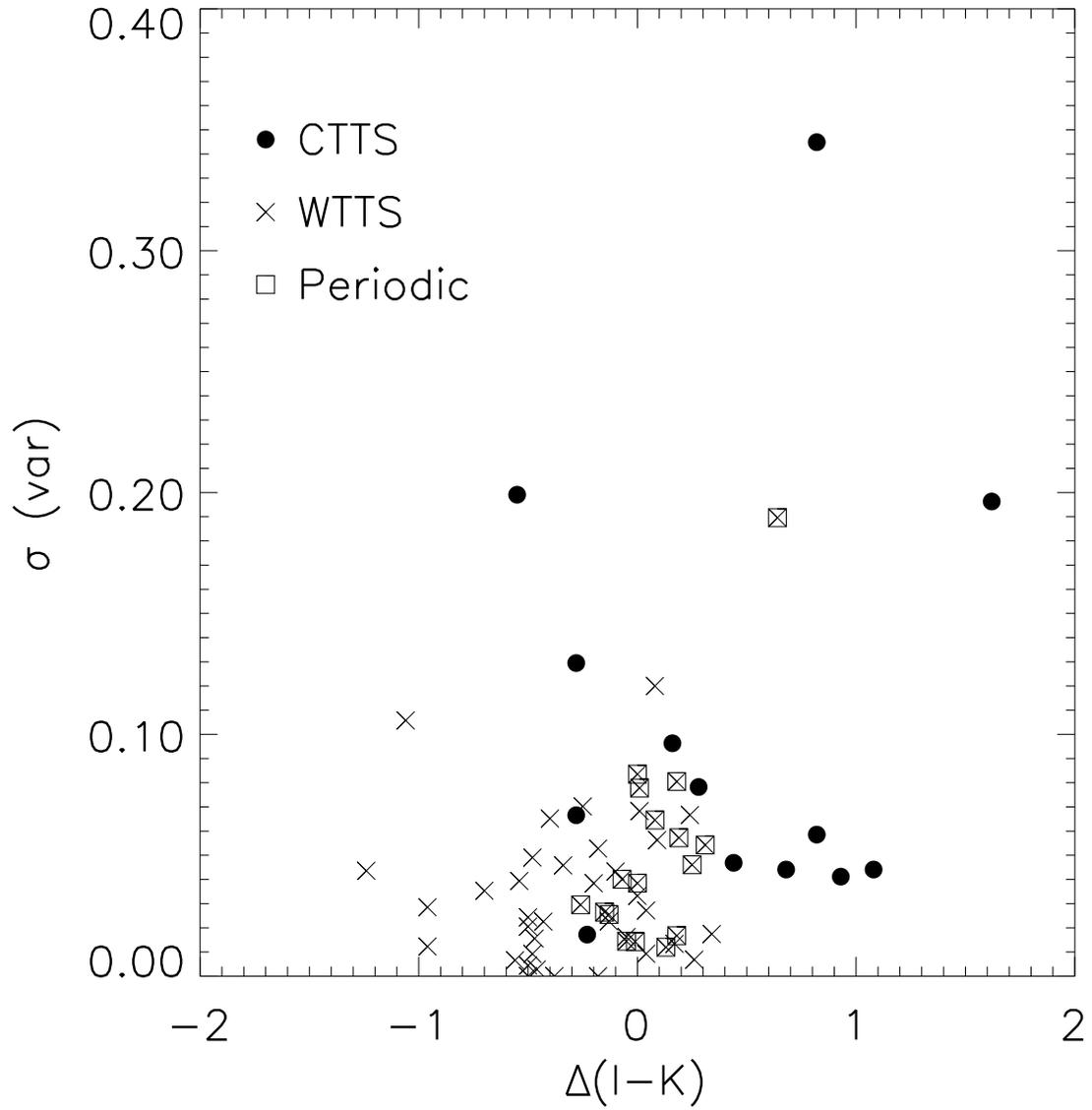} 
\figcaption{Degree of variability, as measured by $\sigma_{var}$, 
versus infrared excess emission, as measured by $\Delta$(I-K).}  
\end{figure}

\end{document}